\documentstyle[12pt,psfig,amsmath,graphicx]{article}
\input amssym.tex
\renewcommand{\baselinestretch}{1.1}

\def\be{\begin{equation}}
\def\ee{\end{equation}}

\def\TL{\hfil$\displaystyle{##}$}

\def\TT{\hbox{##}}
\def\seqalign#1#2{\vcenter{\openup1\jot
  \halign{\strut #1\cr #2 \cr}}}

\def\fixit#1{}
\def\mop#1{\mathop{\rm #1}\nolimits}

\def\Tr{\mop{Tr}}
\def\href#1#2{#2}

\newcommand{\vev}[1]{\langle{#1}\rangle}
\newcommand{\half}{\frac{1}{2}}

\newcommand{\eqn}[2]{\begin{equation} \label{#1} #2 \end{equation}}
\newcommand{\eqalign}[2]{\begin{eqnarray}\label{#1} #2 \end{eqnarray}}

\newcommand{\CO}{{\cal O}}
\newcommand{\CN}{{\cal N}}
\newcommand{\CA}{{\cal A}}

\newcommand{\gb}[1]{\langle#1]}
\newcommand{\rf}[1]{(\ref{#1})}
\def\CM{{\cal M}}
\def\l{\lambda}
\def\tl{\tilde\lambda}
\def\CP{\Bbb{CP}}
\def\RP{\Bbb{RP}}

\begin{document}
\baselineskip=16pt
\pagestyle{plain}
\setcounter{page}{1}

\begin{titlepage}

\begin{flushright}
hep-th/0504194 \\
\end{flushright}
\vfil

\begin{center}
{\huge Lectures on Twistor Strings}\\ \huge{ and} \linebreak
\huge{Perturbative Yang-Mills Theory}
\end{center}

\vfil
\begin{center}
{\large Freddy Cachazo and Peter Svr\v{c}ek\footnote{On leave from
Princeton University.}
}\end{center}

$$\seqalign{\span\TL\, & \sl\span\TT}{
 & Institute for Advanced Study,
 Princeton, NJ  08540, USA  \cr
}$$
\vfil

\begin{center}
{\large Abstract}
\end{center}

\noindent Recently, Witten proposed a topological string theory in
twistor space that is dual to a weakly coupled gauge theory. In
this lectures we will discuss aspects of the twistor string
theory. Along the way we will learn new things about Yang-Mills
scattering amplitudes. The string theory sheds light on Yang-Mills
perturbation theory and leads to new methods for computing
Yang-Mills scattering amplitudes\footnote{These lecture notes are
based on lectures given by the second author at RTN Winter School
on Strings,
 Supergravity and Gauge Theories, 31/1-4/2 2005 SISSA, Trieste, Italy}.

\vfil
\begin{flushleft}
April, 2005
\end{flushleft}
\end{titlepage}
\newpage
\renewcommand{\thefootnote}{\arabic{footnote}}
\setcounter{footnote}{0}
\renewcommand{\baselinestretch}{1.2}  
\tableofcontents
\def\bar{\overline}

\section{Introduction}

The idea that a gauge theory should be dual to a string theory
goes back to 't Hooft \cite{'tHooft:1973jz}.  't Hooft considered
$U(N)$ gauge theory in the large $N$ limit while keeping
$\lambda=g_{YM}^2 N$ fixed. He observed that the perturbative
expansion of Yang-Mills can be reorganized in terms of Riemann
surfaces, which he interpreted as an evidence for a hypothetical
dual string theory with string coupling $g_s\sim 1/N.$

In 1997, Maldacena proposed a concrete example of this duality
\cite{Maldacena:1997re}. He considered the maximally
supersymmetric Yang-Mills theory and conjectured that it is dual
to type IIB string theory on $AdS_5\times S^5.$ This duality led
to many new insights from string theory about gauge theories and
vice versa. At the moment, we have control over the duality only
for strongly coupled gauge theory. This corresponds to the limit
of large radius of $AdS_5\times S^5$ in which the string theory is
well described by supergravity. However, QCD is asymptotically
free, so we would also like to have a string theory description of
a weakly coupled gauge theory.

In weakly coupled field theories, the natural object to study is
the perturbative $S$ matrix. The perturbative expansion of the $S$
matrix is conventionally computed using Feynman rules. Starting
from early studies of de Witt \cite{DeWitt:1967uc}, it was
observed that scattering amplitudes show simplicity that is not
apparent from the Feynman rules. For example, the maximally
helicity violating (MHV) amplitudes can be expressed as  simple
holomorphic functions.

Recently, Witten proposed a string theory that is dual to a weakly
coupled $\CN=4$ gauge theory \cite{Witten:2003nn}. The
perturbative expansion of the gauge theory is related to
D-instanton expansion of the string theory. The string theory in
question is the topological open string B-model on a Calabi-Yau
supermanifold $\CP^{3|4},$ which is a supersymmetric
generalization of  Penrose's twistor space.

At tree level, evaluating the instanton contribution has led to
new insights about scattering amplitudes. `Disconnected'
instantons give the MHV diagram construction of  amplitudes in
terms of Feynman diagrams with vertices that are suitable
off-shell continuations of the MHV amplitudes
\cite{Cachazo:2004kj}. The `connected' instanton contributions
express the amplitudes as integrals over the moduli space of
holomorphic curves in twistor space \cite{Roiban:2004yf}.
Surprisingly, the MHV diagram construction and the connected
instanton integral can be related via localization on the moduli
space \cite{Gukov:2004ei}.

Despite the successes of the twistor string theory at tree level,
there are still many open questions. The most pressing issue is
perhaps the closed strings that give $\CN=4$ conformal
supergravity \cite{Berkovits:2004jj}. At tree level, it is
possible to recover the Yang-Mills scattering amplitudes by
extracting the single-trace amplitudes. At loop level, the single
trace gluon scattering amplitudes receive contributions from
internal supergravity states, so it would be difficult to extract
the Yang-Mills contribution to the gluon scattering amplitudes.
Since, $\CN=4$ Yang-Mills theory is consistent without conformal
supergravity, it is likely that there exists a version of the
twistor string theory that is dual to pure Yang-Mills theory.
Indeed, the MHV diagram construction that at tree level has been
derived from twistor string theory seems to compute loop
amplitudes as well \cite{Brandhuber:2004yw}.

The study of twistor structure of scattering amplitudes has
inspired new developments in perturbative Yang-Mills theory
itself. At tree level, this has led to recursion relations for
on-shell amplitudes \cite{Britto:2004ap}. At one loop, unitarity
techniques \cite{Bern:1994cg,Bern:1994zx} have been used to find
new ways of computing  $\CN=4$ \cite{Britto:2004nc} and $\CN=1$
\cite{Britto:2005ha} Yang-Mills amplitudes.

In these lectures we will discuss aspects of twistor string
theory. Along the way we will learn lessons about Yang-Mills
scattering amplitudes. The string theory sheds light on Yang-Mills
perturbation theory and leads to new methods for computing
Yang-Mills scattering amplitudes. In the last section, we will
describe further developments in perturbative Yang-Mills.

\section{Helicity Amplitudes}
\subsection{Spinors}
\label{spinors}

Recall\footnote{The sections $2-4$
 are based on lectures given by E. Witten at PITP, IAS Summer 2004}
 that the complexified Lorentz group is locally isomorphic
to
\begin{equation}
SO(3,1,\Bbb C)\cong Sl(2,\Bbb C)\times Sl(2,\Bbb C),
\label{sothree}
\end{equation}
hence the finite dimensional representations are classified as
$(p,q)$ where $p$ and $q$ are integer or half-integer. The
negative and positive chirality spinors transform in the
representations $(1/2,0)$ and $(0,1/2)$ respectively. We write
generically $\lambda_a,a=1,2$ for a spinor transforming as
$(1/2,0)$ and $\tilde\lambda_{\dot a}, \dot a=1,2$ for a spinor
transforming as $(0,1/2).$

The spinor indices of type $(1/2,0)$ are raised and lowered using
the antisymmetric tensors $\epsilon_{ab}$ and $\epsilon^{ab}$
obeying $\epsilon_{12}=1$ and
$\epsilon^{ac}\epsilon_{cb}=\delta^a{}_b$ \eqalign{raise}{
\lambda^a = \epsilon^{ab}\lambda_b\qquad \lambda_a =
\epsilon_{ab}\lambda^b.} Given two spinors $\lambda$ and
$\lambda',$ both of negative chirality, we can form the Lorentz
invariant product
\eqn{skew}{\vev{\lambda,\lambda'}=\epsilon_{ab}\lambda^a\lambda'{}^b.}
It follows that $\vev{\lambda,\lambda'}=-\vev{\lambda',\lambda}$,
so the product is antisymmetric in its two variables. In
particular, $\vev{\lambda,\lambda'}=0$ implies that $\lambda$
equals $\lambda'$  up to a scaling $\lambda^a=c \lambda'^a.$

Similarly, we lower and raise the indices of positive chirality
spinors with the antisymmetric tensor $\epsilon_{\dot a\dot b}$
and its inverse $\epsilon^{\dot a \dot b}.$ For two spinors
$\tilde\lambda$ and $\tilde\lambda',$ both of positive chirality
we define the antisymmetric product
\eqn{angled}{[\tilde\lambda,\tilde\lambda']=-[\tilde\lambda',\tilde\lambda]=\epsilon_{\dot
a\dot b}\tilde\lambda^{\dot a}\tilde\lambda'^{\dot b}.}

The vector representation of $SO(3,1,\Bbb C)$ is the $(1/2,1/2)$
representation. Thus a momentum vector $p_\mu, \mu=0,\dots,3$ can
be represented as a bi-spinor $p_{a\dot a}$ with one spinor index
$a$ and $\dot a$ of each chirality. The explicit mapping from
$p_\mu$ to $p_{a \dot a}$ can be made using the chiral part of the
Dirac matrices.  In signature $+---,$ one can take the Dirac
matrices to be \begin{equation}\gamma^\mu=\begin{pmatrix}0 &
\sigma^\mu \cr \bar\sigma^\mu &0\end{pmatrix},\end{equation} where
$\sigma^\mu=(1,\vec\sigma), \bar\sigma^\mu=(1,-\vec\sigma)$ with
$\vec\sigma$ being the $2\times 2$ Pauli matrices.  For any
vector, the relation between  $p_\mu,$ and $p_{a\dot a}$ is
\eqn{transf}{ p_{a\dot a}= p_\mu \sigma^\mu_{a \dot
a}=p_0+\vec\sigma\cdot\vec p.} It follows that, \eqn{detr}{p_\mu
p^\mu= \det(p_{a\dot a}).} Hence, $p_\mu$ is lightlike if the
corresponding determinant is zero. This is equivalent to the rank
of the $2\times2$ matrix $p_{a\dot a}$ being less than or equal to
one. So $p^\mu$ is lightlike precisely, when it can be written as
a product \eqn{bispi}{p_{a\dot a}=\lambda_a\tilde\lambda_{\dot a}}
for some  spinors $\lambda_a$ and $\tilde\lambda_{\dot a}.$ For a
given null vector $p,$ the spinors $\lambda$ and $\tilde\lambda$
are unique up to a scaling
\eqn{scalm}{(\lambda,\tilde\lambda)\rightarrow
(t\lambda,t^{-1}\tilde\lambda)\qquad t\in \Bbb C^\ast.} There is
no continuous way to pick $\lambda$ as a function $p.$ In
Minkowski signature, the $\lambda$'s form the Hopf line bundle
over the sphere $S^2$ of directions of the lightlike vector $p.$

For complex momenta, the spinors $\lambda^a$ and
$\tilde\lambda^{\dot a}$ are independent complex variables, each
of which parameterizes a copy of $\Bbb  {CP}^1.$ Hence, the
complex lightcone $p_\mu p^\mu=0$ is a complex cone over the
connected manifold $\Bbb {CP}^1\times \Bbb {CP}^1.$

For real null momenta  in Minkowski signature $+---$, we can fix
the scaling up to a $Z_2$ by requiring $\lambda^a$ and
$\tilde\lambda^{\dot a}$ to be complex conjugates
\eqn{fixe}{\bar\lambda^{\dot a}=\pm\tilde\lambda^{\dot a}.} Hence,
the negative chirality spinors $\lambda$ are conventionally called
`holomorphic' and the positive chirality spinor
`anti-holomorphic.' In \rf{fixe} the $+$ sign is for a future
pointing null vector $p^\mu,$ and $-$ is for a past pointing
$p^\mu.$

One can also consider other signatures. For example in the
signature $++--,$  the spinors $\lambda$ and $\tilde\lambda$ are
real and independent. Indeed, with signature $++--,$ the Lorentz
group is $SO(2,2),$ which is locally isomorphic to $Sl(2,\Bbb
R)\times Sl(2,\Bbb R).$ Hence, the spinor representations are
real.

Let us remark, that if $p$ and $p'$ are two lightlike vectors
given by $p_{a\dot a}=\lambda_a\tilde\lambda_{\dot a}$ and
$p'_{a\dot a}=\lambda'_a\tilde\lambda'_{\dot a}$ then their scalar
product can be expressed as\footnote{This differs from the `-'
sign convention used in the perturbative QCD literature.}
\eqn{product}{2 p\cdot
p'=\vev{\lambda,\lambda'}[\tilde\lambda,\tilde\lambda'].}

Given $p,$ the additional physical information in $\lambda$ is
equivalent to a choice of wavefunction of a helicity $-1/2$
massless particle with momentum $p.$ To see this, we write the
chiral Dirac equation for a negative chirality spinor $\psi^a$
\eqn{chdir}{0=i\sigma^\mu_{a\dot a} \partial_\mu \psi^a.} A plane
wave $\psi^a=\rho^a\exp(ip\cdot x)$ satisfies this equation if and
only if $p_{a\dot a}\rho^a=0.$ Writing  $p_{a\dot a}=\lambda_a
\tilde\lambda_{\dot a},$ we get $\lambda_a\rho^a=0,$ that is
$\rho^a= c\cdot \lambda^a$ for a constant $c.$ Hence the negative
helicity fermion has wavefunction \eqn{negw}{\psi^a=c\lambda^a
\exp(ix_{a\dot a} \lambda^a\tilde\lambda^{\dot a}).} Similarly,
$\tilde\lambda$ defines a wavefunction for a helicity $+1/2$
fermion $\psi^{\dot a}=c \tilde\lambda^{\dot a}\exp(ix_{a\dot a}
\lambda^a\tilde\lambda^{\dot a}).$

There is an analogous description of wavefunctions of massless
particles of helicity $\pm1.$ Usually, we describe massless gluons
with their momentum vector $p^\mu$ and polarization vector
$\epsilon^\mu.$ The polarization vector obeys the constraint
\eqn{decop}{p_\mu\ \epsilon^\mu=0} that represents the decoupling
of longitudinal modes and it is subject to the gauge invariance
\eqn{gaugi}{\epsilon^\mu\rightarrow\epsilon^\mu+wp^\mu,} for any
constant $w.$ Suppose that instead of being given only a lightlike
vector $p_{a\dot a}$, one is also given a decomposition $p_{a\dot
a}=\lambda_a\tilde\lambda_{\dot a}.$ Then we have enough
information to determine the polarization vector up to a gauge
transformation once the helicity is specified. For a positive
helicity gluon, we take \eqn{pospol}{\epsilon_{a\dot
a}^+=\frac{\mu_a\tilde\lambda_{\dot a}}{\vev{\mu,\lambda}},} where
$\mu$ is any negative chirality spinor that is not a multiple of
$\lambda.$ To get a negative helicity polarization vector, we take
\eqn{negpol}{\epsilon_{a\dot a}^-=\frac{\lambda_a\tilde\mu_{\dot
a}}{[\tilde\lambda,\tilde\mu]},} where $\tilde\mu$ is any positive
chirality spinor that is not a multiple of $\tilde\lambda.$ We
will explain the expression for the positive helicity vector. The
negative helicity case is similar.

Clearly, the constraint \eqn{con}{p^\mu\ \epsilon^+_\mu=p^{a\dot
a}\epsilon^+_{a\dot a}=0}  holds because $\tilde\lambda^{\dot
a}\tilde\lambda_{\dot a}=0.$ Moreover, $\epsilon^+$ is also
independent of $\mu^a$ up to a gauge transformation. To see this,
notice that $\mu$ lives in a two dimensional space that is spanned
with $\lambda$ and $\mu.$ Hence, any change in $\tilde\mu$ is of
the form \eqn{vam}{\delta \mu^a=\alpha\mu^a+\beta\l^a} for some
parameters $\alpha$ and $\beta.$ The polarization vector
\rf{pospol} is invariant under the $\alpha$ term, because this
simply rescales $\mu$ and $\epsilon^+_{a \dot a}$ is invariant
under the rescaling of $\mu.$ The $\beta$ term amounts to a  gauge
transformation of the polarization vector
\eqn{poch}{\delta\epsilon^+_{a\dot a}=
\beta{\lambda_a\tilde\lambda_{\dot a}\over \vev{\mu,\lambda}}.}

Under the scaling $(\lambda,\tilde\lambda)\rightarrow
(t\lambda,t^{-1}\tilde\lambda),\, t\in {\Bbb C}^\ast$ the
polarization vectors scale like \eqn{scal}{\epsilon^-\rightarrow
t^{+2}\epsilon^- \qquad \epsilon^+\rightarrow t^{-2}\epsilon^+.}
This could have been anticipated, since $\tilde\lambda_{\dot a}$
gives the wavefunction of a helicity $+1/2$ particle  so a
helicity $+1$ polarization vector should scale like
$\tilde\lambda^2.$ Similarly, the helicity $-1$ polarization
vector scales like $\lambda^2.$

To show  more directly that $\epsilon^+$ describes a massless
particle of helicity $+1,$ we must show that the corresponding
linearized field strength $F_{\mu\nu}=\partial_\mu
A_\nu-\partial_\nu A_\mu$ is anti-selfdual. Indeed, the field
strength written in a bispinor notation has the decomposition
\eqn{sef}{F_{a\dot a b\dot b}=\epsilon_{ab}\tilde f_{\dot a\dot
b}+\epsilon_{\dot a \dot b}f_{ab},} where $f_{ab}$ and $\tilde
f_{\dot a \dot b}$ are the selfdual and anti-selfdual parts of
$F.$ Substituting $A_{a\dot a}=\epsilon_{a\dot a}^+\exp(ix_{a\dot
a}\lambda^a\tl^{\dot a})$ we find that $F_{a\dot a b\dot b}=
\epsilon_{ab} \tl_{\dot a}\tl_{\dot b}\exp(ix_{a\dot a}
\lambda^a\tl^{\dot a}).$

So far, we have seen that the wavefunction of a massless particle
with helicity $h$ scales under
$(\lambda,\tilde\lambda)\rightarrow(t\lambda,t^{-1}\tilde\lambda)$
as $t^{-2h}$ if $|h|\leq1.$ This is true for any $h,$ as can be
seen from the following argument. Consider a massless particle
moving in the $\vec n$ direction. Then a rotation by angle
$\theta$ around the $\vec n$ axis acts on the spinors as
\eqn{rota}{(\lambda,\tilde\lambda)\rightarrow
(e^{-i\theta/2}\lambda,e^{+i\theta/2}\tilde\lambda).} Hence,
$\lambda,\tilde\lambda$ carry $-\half$ or $+\half$ units of
angular momentum around the $\vec n$ axis. Clearly, a massless
particle of helicity $h$ carries $h$ units of angular momentum
around the $\vec n$ axis. Hence the wavefunction of the particle
gets transformed as $\psi\rightarrow e^{ih\theta}\psi$ under the
rotation around $\vec n$ axis, so it obeys the auxiliary condition
\eqn{scali}{\left(\lambda^a\frac{\partial}{\partial\lambda^a}-\tilde\lambda^{\dot
a}\frac{\partial}{\partial\tilde\lambda^{\dot a}}\right)
\psi(\lambda,\tilde\lambda)=-2h\psi(\lambda,\tilde\lambda).}
Clearly, this constraint holds for wavefunctions of massless
particles of any spin. The spinors $\lambda,\tilde\lambda$ give us
a convenient way of writing the wavefunction of massless particle
of any spin, as we have seen in detail above for particles with
$|h|\leq 1.$

\subsection{Scattering Amplitudes}

Let us consider scattering  of massless particles in four
dimensions. Consider the situation with $n$ particles of momenta
$p_1,p_2,\dots, p_n.$ For scattering of scalar particles, the
initial and final states of the particles are completely
determined by their momenta. The scattering amplitude is simply a
function of the momenta $p_i,$
\eqn{ascal}{A_{scalar}(p_1,p_2,\dots,p_n).} In fact, by Lorentz
invariance, it is a function of the Lorentz invariant products
$p_i\cdot p_j$ only.

For particles with spin, the scattering amplitude is a function of
both the momenta $p_i$ and the wavefunctions $\psi_i$ of each
particle \eqn{ascat}{A(p_1,\psi_1;\dots;p_n,\psi_n).} Here, $A$ is
linear in each of the wavefunctions $\psi_i.$ The description of
$\psi_i$ depends on the spin of the particle. As we have seen
explicitly above in the case of massless particles of spin $\half$
or $1,$ the spinors $\lambda,\tilde\lambda$ give a unified
description of the wavefunctions of particles with spin. Hence, to
describe the wavefunctions, we specify for each particle the
helicity $h_i$ and the spinors $\lambda_i$ and $\tilde\lambda_i.$
The spinors determine the momenta $p_i=\lambda_i\tilde\lambda_i$
and the wavefunctions $\psi_i(\lambda_i,\tilde\lambda_i, h_i)$. So
for massless particles with spin, the scattering amplitude is a
function of the spinors and helicities of the external particles
\eqn{scaf}{A(\lambda_1,\tilde\lambda_1,h_1;\dots;\lambda_n,\tilde\lambda_n,h_n).}
In labelling the helicities we take all particles to be incoming.
To obtain an amplitude with incoming particles as well as outgoing
particles, we use crossing symmetry, that relates an incoming
particle of one helicity  to an outgoing particle of the opposite
helicity.

It follows from \rf{scali} that the amplitude obeys the conditions
\eqn{scalia}{\left(\lambda^a_i\frac{\partial}{\partial\lambda^a_i}-\tilde\lambda_i^{\dot
a}\frac{\partial}{\partial\tilde\lambda^{\dot a}_i}\right)
A(\lambda_i,\tilde\lambda_i,h_i)=-2h_iA(\lambda_i,\tilde\lambda_i,h_i)}
for each particle $i,$ with helicity $h_i.$ In summary, a general
scattering amplitude of massless particles can be written as
\eqn{scatmas}{A=(2\pi)^4 \delta^4\left(\sum_i
\lambda_i^a\tilde\lambda^{\dot a}_i\right)
A(\lambda_i,\tilde\lambda_i,h_i),} where we have written
explicitly the delta function of momentum conservation.

\begin{figure}
 \centering
 \includegraphics[height=2.0in]{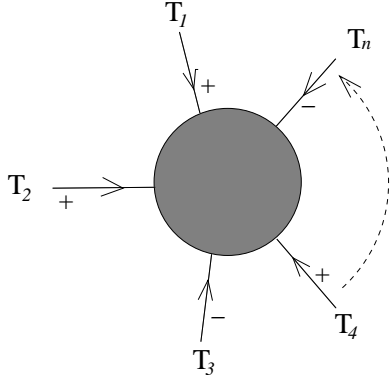}
 \caption{A scattering amplitude of $n$ gluons in Yang-Mills theory.
          Each gluon comes with the color factor $T_i,$ spinors
          $\lambda_i,\tilde\lambda_i$ and helicity label $h_i=\pm1.$}
 \label{ampl_notation}
\end{figure}


\subsection{Maximally Helicity Violating Amplitudes}

To make the discussion more concrete, we consider tree level
scattering of $n$ gluons in Yang-Mills theory. These amplitudes
are of phenomenological importance. The multijet production at LHC
will be dominated by tree level QCD scattering.

Consider Yang-Mills theory with gauge group $U(N).$ Recall that
tree level scattering amplitudes are planar and lead to single
trace interactions. In an index loop, the gluons are attached in a
definite cyclic order, say $1,2,\dots, n.$ Then the amplitude
comes with a group theory factor $\Tr T_1T_2\dots T_n.$ It is
sufficient to give the amplitude with one cyclic order. The full
amplitude is obtained from this by summing over the cyclic
permutations to achieve Bose symmetry
\eqn{boses}{A=g^{n-2}(2\pi)^4\delta^4\left(\sum_i^n
p_i\right)\CA(1,2,\dots, n)\Tr (T_1T_2\dots T_n)\, +\,
permutations.} Here, $g$ is the coupling constant of the gauge
theory. In the rest of the lecture notes, we will always consider
gluons in the cyclic order $1,2,\dots, n$ and we will omit the
group theory factor and the delta function of momentum
conservation in writing the formulas. Hence we will consider the
`reduced color ordered amplitude' $\CA(1,2,\dots, n).$

The scattering amplitude with $n$ incoming gluons of the same
helicity vanishes. So does the amplitude, for $n\geq 3,$ with
$n-1$ incoming gluons of one helicity and one of the opposite
helicity. The first nonzero amplitudes, the maximally helicity
violating (MHV) amplitudes, have $n-2$ gluons of one helicity and
two gluons of the other helicity. Suppose that gluons $r,s$ have
negative helicity and the rest of gluons have positive helicity.
Then the tree level amplitude, stripped of the momentum delta
function and the group theory factor, is
\eqn{mhv}{\CA(r^-,s^-)=g^{n-2}\frac{\vev{\lambda_r,\lambda_s}^4}{\prod_{k=1}^n
\vev{\lambda_k,\lambda_{k+1}}}.} The amplitude $\CA(r^+,s^+)$ with
gluons $r,s$ of positive helicity and the rest of the gluons of
negative helicity follows from \rf{mhv} by exchange
$\vev{}\leftrightarrow [].$ Note, that the amplitude has the
correct homogeneity in each variable. It is homogeneous of degree
$-2$ in $\lambda_i$ for positive helicity gluons; and of degree
$-2$ for negative helicity gluons $i=r,s$ as required by the
auxiliary condition \rf{scalia}. The amplitude $\CA$ is sometimes
called `holomorphic' because it depends on the `holomorphic'
spinors $\lambda_i$ only.

\section{Twistor Space}

\subsection{Conformal Invariance of Scattering Amplitudes}

Before discussing twistor space, let us show the conformal
invariance of the MHV tree level amplitude. Firstly, we need to
construct representation of the conformal group generators in
terms of the spinors $\l,\tl.$ We will consider the conformal
generators for a single particle. The generators of the
$n$-particle system are given by the sum of the generators over
the $n$ particles.

Some of the generators are clear. The Lorentz generators are the
first order differential operators
\eqalign{lorentz}{J_{ab}&=&\frac{i}{2}\left(\lambda_a
\frac{\partial}{\partial\lambda^b}+\lambda_b\frac{\partial}{\partial\lambda^a}
\right)\cr  \tilde J_{\dot a \dot
b}&=&\frac{i}{2}\left(\tilde\lambda_{\dot
a}\frac{\partial}{\partial\tilde\lambda^{\dot
b}}+\tilde\lambda_{\dot
b}\frac{\partial}{\partial\tilde\lambda^{\dot a}}\right).} The
momentum operator is the multiplication operator
\eqn{momentum}{P_{a\dot a}=\lambda_a\tilde\lambda_{\dot a}.} The
remaining generators are the dilatation operator $D$ and the
generator of special conformal transformation $K_{a\dot a}.$ The
commutation relations of the dilatation operator are
\eqn{comd}{[D,P]=iP,\qquad [D,K]=-iK,} so $P$ has dimension $+1$
and $K$ has dimension $-1.$ We see from (\ref{momentum}) that it
is natural to take $\lambda$ and $\tilde\lambda$ to have dimension
$1/2.$ Hence, a natural guess for the special conformal generator
respecting all the symmetries is \eqn{kguess}{K_{a\dot
a}=\frac{\partial^2}{\partial\lambda^a\partial\tilde\lambda^{\dot
a}}.} We find the dilatation operator $D$ from the closure of the
conformal algebra.  The commutation relation
\eqn{comrel}{[K_{a\dot a},P^{b\dot b}]=-i\left(\delta_a{}^b\tilde
J_{\dot a}{}^{\dot b}+\delta_{\dot a}{}^{\dot
b}J_{a}{}^b+\delta_a{}^b\delta_{\dot a}{}^{\dot b}D\right)}
determines the dilatation operator to be
\eqn{dilat}{D=\frac{i}{2}\left(\lambda^a\frac{\partial}{\partial\lambda^a}
+\tilde\lambda^{\dot a}\frac{\partial}{\partial\tilde\lambda^{\dot
a}}+2\right).}

We are now ready to verify that the MHV amplitude
\eqn{mhvdil}{\CA(r^-,s^-)=(2\pi)^4\delta^4\left(\sum_i\lambda_i^a\tilde\lambda^{\dot
a}_i\right)
\frac{\vev{\lambda_r,\lambda_s}^4}{\prod_{k=1}^n\vev{\lambda_k,\lambda_{k+1}}},}
is invariant under the conformal group. The Lorentz generators are
clearly symmetries of the amplitude. The momentum operator
annihilates the amplitude thanks to the delta function of momentum
conservation.

It remains to verify that the amplitude is annihilated by $D$ and
$K.$ For simplicity, we will only consider the dilatation operator
$D.$ The numerator contains the delta function of momentum
conservation which has dimension $D=-4$ and the factor
$\vev{\lambda_r,\lambda_s}^4$ of dimension $4.$ Hence, $D$
commutes with the numerator. So we are left with the denominator
\eqn{denomi}{\frac{1}{\prod_{k=1}^n\vev{\lambda_k,\lambda_{k+1}}}.}
This is annihilated by $D_k$ for each particle $k,$ since the $-2$
coming from the second power of $\lambda_k$ in the denominator
gets cancelled against the $+2$ from the definition of the
dilatation operator \rf{dilat}.

\subsection{Transform to Twistor Space}
\label{transform}

We have demonstrated conformal invariance of the MHV amplitude,
however the representation of the conformal group that we have
encountered above is quite exotic. The Lorentz generators are
first order differential operators, but the momentum is a
multiplication operator and the special conformal generator is a
second order differential operator.

We can put the action of the conformal group into a more standard
form if we make the following transformation
\eqalign{four}{\tilde\lambda_{\dot a}&\rightarrow
i\frac{\partial}{\partial\mu^{\dot a}} \cr
\frac{\partial}{\partial\tilde\lambda^{\dot a}}&\rightarrow
i\mu_{\dot a}.}  Making this substitution we have arbitrarily
chosen to Fourier transform $\tl$ rather than $\l.$ This choice
breaks the symmetry between positive and negative helicities. The
amplitude with $n_1$ positive helicity and $n_2$ negative helicity
gluons has different description in twistor space from an
amplitude with $n_2$ positive helicity gluons and $n_1$ negative
helicity gluons.

Upon making this substitution, all operators become first order.
The Lorentz generators take the form \eqalign{lorentztwo}{J_{ab}
&=&\frac{i}{2}\left(\lambda_a
\frac{\partial}{\partial\lambda^b}+\lambda_b\frac{\partial}{\partial\lambda^a}
\right)\cr  \tilde J_{\dot a \dot b} &=&\frac{i}{2}\left(\mu_{\dot
a}\frac{\partial}{\partial\mu^{\dot b}}+\mu_{\dot
b}\frac{\partial}{\partial\mu^{\dot a}}\right).} The momentum and
special conformal operators become \eqalign{fir}{P_{a\dot
a}&=&i\lambda_a\frac{\partial}{\partial\mu^{\dot a}} \cr K_{a\dot
a}&=&i\mu_{\dot a}\frac{\partial}{\partial\lambda^a}.} Finally,
the dilatation operator (\ref{dilat}) becomes a {\it homogeneous}
first order operator
\eqn{dilafirst}{D=\frac{i}{2}\left(\lambda^a{\partial\over
\partial\lambda^a} -\mu^{\dot a}\frac{\partial}{\partial\mu^{\dot
a}}\right).}

This representation of the four dimensional conformal group is
easy to explain. The conformal group of Minkowski space is
$SO(4,2)$ which is the same as $SU(2,2).$ $SU(2,2),$ or its
complexification $Sl(4,\Bbb C),$ has an obvious four-dimensional
representation acting on \eqn{zi}{Z^I=(\lambda^a,\mu^{\dot a}).}
$Z^I$ is called a twistor and the space $\Bbb C^4$ spanned by
$Z^I$ is called the twistor space. The action of $Sl(4,\Bbb C)$ on
the $Z^I$ is generated by $15$ traceless matrices $\Lambda^I{}_J,\
I,J=1,\dots, 4,$ that correspond to the $15$ first order operators
$J_{ab}, \tilde J_{\dot a \dot b}, D, P_{a\dot a}, K_{a\dot a}.$

If we are in signature $++--,$ the conformal group is $SO(3,3)
\cong Sl(4,\Bbb R).$ The twistor space is a copy of $\Bbb R^4$ and
we can consider $\lambda$ and $\mu$ to be real. In the Euclidean
signature $++++$, the conformal group is $SO(5,1)\cong SU^\ast(4)$
where $SU^\ast(4)$ is the noncompact version of $SU(4),$ so we
must think of twistor space as a copy of $\Bbb C^4.$

For signature $++--,$ where $\tilde\lambda$ is real, the
transformation from momentum space scattering amplitudes to
twistor space scattering amplitudes is made by a simple Fourier
transform that is familiar from quantum mechanics
\eqn{fouriert}{\tilde \CA(\lambda_i,\mu_i)= \int\prod_{j=1}^n
{d^2\tilde\lambda_j\over (2\pi)^2} \exp(i[\mu_j,\tilde\lambda_j])
\CA(\lambda_i,\tilde\lambda_i).} The same Fourier transform turns
a momentum space wavefunction $\psi(\lambda,\tilde\lambda)$ to a
twistor space wavefunction
\eqn{wavefo}{\tilde\psi(\lambda,\mu)=\int {d^2\tilde\lambda\over
(2\pi)^2} \exp(i[\mu,\tilde\lambda]) \psi(\lambda,\tilde\lambda).}

Recall that the scattering amplitude of  massless particles obeys
the auxiliary condition
\eqn{sara}{\left(\lambda^a_i\frac{\partial}{\partial\lambda^a_i}-\tilde\lambda_i^{\dot
a}\frac{\partial}{\partial\tilde\lambda^{\dot a}_i}\right)
\CA(\lambda_i,\tilde\lambda_i,h_i)=-2h_i\CA(\lambda_i,\tilde\lambda_i,h_i)}
for each particle $i,$ with helicity $h_i.$ In terms of
$\lambda_i$ and $\mu_i,$ this becomes
\eqn{lica}{\left(\lambda^a_i\frac{\partial}{\partial\lambda^a_i}+\mu_i^{\dot
a}\frac{\partial}{\partial\mu^{\dot a}_i}\right)
\tilde\CA(\lambda_i,\mu_i,h_i)=-(2h_i+2)\tilde\CA(\lambda_i,\mu_i,h_i).}
There is a similar condition for the twistor wavefunctions of
particles. The operator on the left hand side coincides with
$Z^I{\partial\over
\partial Z^I}$ that generates the scaling of the twistor coordinates
\eqn{rescal}{Z^I\rightarrow t Z^I, \qquad t\in \Bbb C^\ast.}

So the wavefunctions and scattering amplitudes have known behavior
under the $\Bbb C^\ast$ action $Z^I\rightarrow t Z^I.$ Hence, we
can identify the sets of $Z^I$ that differ by the scaling
$Z^I\rightarrow t Z^I$ and throw away the point $Z^I=0.$ We get
the projective twistor space\footnote{The twistor wavefunctions
\rf{wavefo} are regular only on the subset $\CP'^{3|4}$ of
$\CP^{3|4}$ with $(\lambda^1,\lambda^2)\neq(0,0),$ which is the
precise definition of the projective twistor space. In the rest of
the lecture notes, we do not distinguish between these two spaces,
unless necessary.} $\Bbb {CP}^3$ or $\Bbb {RP}^3$ if $Z^I$ are
complex or real-valued. The $Z^I$ are the homogeneous coordinates
on the projective twistor space. It follows from \rf{lica} that,
the scattering amplitudes are homogeneous functions of degree
$-2h_i-2$ in the twistor coordinates $Z_i^I$ of each particle
particle. In the complex case, this means that scattering
amplitudes are sections of the complex line bundle ${\cal
O}(-2h_i-2)$ over a $\Bbb {CP}^3_i$ for each particle. For further
details on twistor transform, see any standard textbook on twistor
theory, e.g. \cite{Huggett:1986fs,Atiyah:1979iu}.

\subsection{Scattering Amplitudes in Twistor Space}
\label{twist}

In an $n$ gluon scattering process, after the Fourier transform
into twistor space, the external gluons are associated with points
$P_i$ in the projective twistor space. The scattering amplitudes
are functions of the twistors $P_i,$ that is, they are functions
defined on the product of $n$ copies of twistor space, one for
each particle.

Let us see what happens to the tree level MHV amplitude with $n-2$
gluons of positive helicity and $2$ gluons of negative helicity,
after Fourier transform into twistor space. We work in $++--$
signature, for which the twistor space is a copy of $\Bbb {RP}^3.$
The advantage of $++--$ signature is that the transform to twistor
space is an ordinary Fourier transform and the scattering
amplitudes are ordinary functions on a product of $\Bbb {RP}^3$'s,
one for each particle. With other signatures, the twistor
transform involves $\bar\partial$-cohomology and other
mathematical machinery.

We recall that the MHV amplitude with negative helicity gluons
$r,s$ is
\eqn{mhvrecall}{A(\lambda_i,\tilde\lambda_i)=(2\pi)^4\delta^4(\sum_i
\lambda_i\tilde\lambda_i) f(\lambda_i),} where
\eqn{fln}{f(\lambda_i)=g^{n-2}{\vev{\lambda_r,\lambda_s}^4\over
\prod_k\vev{\lambda_k, \lambda_{k+1}}}.}  The only property of
$f(\lambda_i),$ that we need is that it is a function of the
holomorphic spinors $\lambda_i$ only. It does not depend on the
anti-holomorphic spinors $\tilde\lambda_i.$

We express the delta function of momentum conservation as an
integral \eqn{deltan}{(2\pi)^4\delta^4(\sum_i
\lambda_i^a\tilde\lambda_i^{\dot a})=\int d^4x^{a\dot a}
\exp\left(i x_{b\dot b}\sum_i \lambda_i^b\tilde\lambda_i^{\dot
b}\right).} Hence, we can rewrite the amplitude as \eqn{mhvnew}{
A(\lambda_i,\tilde\lambda_i)=\int d^4x \exp\left(i x_{b\dot
b}\sum_i \lambda_i^b\tilde\lambda_i^{\dot b}\right) f(\lambda_i).}
To transform the amplitude into twistor space, we simply carry out
a Fourier transform with respect to all $\tilde\lambda$'s. Hence,
the twistor space amplitude is
\eqn{mhvint}{A(\lambda_i,\mu_i)=\int
{d^2\tilde\lambda_1\over(2\pi)^2}\dots{d^2\tilde\lambda_n\over
(2\pi)^2}\exp\left(i\sum_{j=1}^n \mu_{j\dot
a}\tilde\lambda_j^{\dot a}\right)\int d^4x\, \exp\left(i x_{b\dot
b}\sum_j \lambda_j^b\tilde\lambda_j^{\dot b}\right) f(\lambda_i).}
The only dependece on $\tl_i$ is in the exponential factors. Hence
the integrals over $\tilde\lambda_j$ gives a product of delta
functions with the result \cite{Nair:1988bq}
\eqn{mhvtwistor}{A(\lambda_i,\mu_i)=\int d^4x \prod_{j=1}^n
\delta^2(\mu_{j\dot a}+x_{a\dot a}\lambda^a_j)f(\lambda_i).} This
equation has a simple geometrical interpretation. Pick some
$x^{a\dot a}$ and consider the equation \eqn{inci}{\mu_{\dot
a}+x_{a\dot a}\lambda^a=0.} The solution set for $x=0$ is a $\Bbb
{RP}^1$ or $\Bbb {CP}^1$ depending on whether the variables are
real or complex. This is true for any $x$ as the equation lets us
solve for $\mu_{\dot a}$ in terms of $\lambda^a.$ So
$(\lambda^1,\lambda^2)$ are the homogeneous coordinates on the
curve.

\begin{figure}
 \centering
 \includegraphics[height=1.5in]{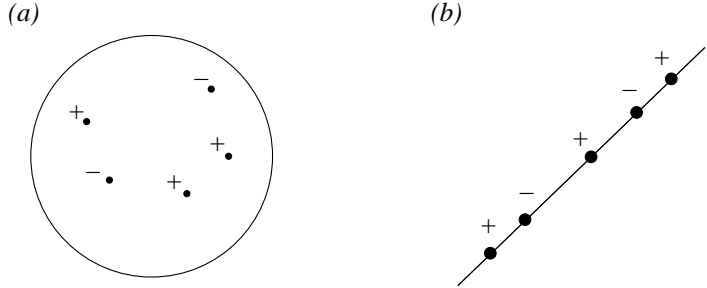}
 \caption{(a) In
          complex twistor space $\CP^3$, the MHV amplitude localizes to a $\Bbb{CP}^1.$
          (b) In the real case, the amplitude is associated to a real line in $\Bbb R^3.$}
 \label{cp_one}
\end{figure}

In real twistor space, which is appropriate for signature $++--,$
the curve $\Bbb {RP}^1$ can be described more intuitively as a
straight line, see fig. \ref{cp_one}. Indeed, throwing away the
set $Z^1=0,$ we can describe the rest of $\Bbb {RP}^3$ as a copy
of $\Bbb R^3$ with the coordinates $x_i=Z_i/Z_1, i=2,3,4.$ The
equations \rf{inci} determine two planes whose intersection is the
straight line in question.

In complex twistor space, the genus zero curve $\Bbb {CP}^1$ is
topologically a sphere $S^2.$ The $\Bbb {CP}^1$  is an example of
a holomorphic curve in $\Bbb {CP}^3.$ The simplest holomorphic
curves are defined by vanishing of a pair of homogeneous
polynomials in the $Z^I$ \eqalign{completeint}{f(Z^1,\dots,
Z^4)&=&0 \cr g(Z^1,\dots, Z^4)&=&0.} If $f$ is  homogeneous of
degree $d_1$ and $g$ is homogeneous of degree $d_2,$ the curve has
degree $d_1 d_2.$ The equations \eqn{incide}{\mu_{\dot b}+x_{b\dot
b}\lambda^b=0, \quad \dot b=1,2} are  both  linear, $d_1=d_2=1$.
Hence the degree of the $\Bbb {CP}^1$ is $d=d_1d_2=1.$ Moreover,
every degree one genus zero curve in $\Bbb {CP}^3$ is of the form
\rf{incide} for some $x^{b\dot b}.$

The area of a holomorphic curve of degree $d,$ using the natural
metric on $\Bbb {CP}^3,$ is $2\pi d.$ So the curves we found with
$d=1$ have the minimal area among nontrivial holomorphic curves.
They are associated with the minimal nonzero Yang-Mills tree
amplitudes, the MHV amplitudes.

Going back to the amplitude \rf{mhvtwistor}, the
$\delta$-functions mean that the amplitude vanishes unless
$\mu_{j\dot a}+x_{a\dot a}\lambda^a_j=0, j=1,\dots n,$ that is,
unless some curve of degree one determined by $x_{a\dot a}$
contains all $n$ points $(\lambda_j,\mu_j).$ The result is that
the MHV amplitudes are supported on genus zero curves of degree
one. This is a consequence of the holomorphy of these amplitudes.

The general conjecture is that an $l$-loop amplitude with $p$
gluons of positive helicity and $q$ gluons of negative helicity is
supported on a holomorphic curve in twistor space. The degree of
the curve is determined by  \eqn{dec}{d=q-1+l.} The genus of the
curve is bounded by the number of the loops \eqn{lops}{g\leq l.}
The MHV amplitudes are a special case of this for $q=2,l=0.$
Indeed the conjecture in this case gives that MHV amplitudes are
supported in twistor space on a genus zero curve of degree one.

The natural interpretation of this is that the curve is the
worldsheet of a string. In some way of describing the perturbative
gauge theory, the amplitudes arise from coupling of the gluons to
a string. In the next two sections we discuss a proposal for such
a string theory due to Witten \cite{Witten:2003nn} in which the
strings in questions are D1-strings. There is an alternative
version of twistor string theory due to Berkovits
\cite{Berkovits:2004hg, Berkovits:2004tx}, discussed in section
\ref{berko}, in which the curves come from fundamental strings.
The Berkovits's twistor string theory seems to give an equivalent
description of the scattering amplitudes. Further proposals
\cite{Neitzke:2004pf,Aganagic:2004yh,Bars:2004dg} have not yet
been used for computing scattering amplitudes.


\section{Twistor String Theory}

In this section, we will describe a string theory that gives a
natural framework for understanding the twistor properties of
scattering amplitudes discussed in previous section. This is a
topological string theory whose target space is a supersymmetric
version of the twistor space.

\subsection{Brief Review of Topological Strings}

Firstly, let us consider an ${\cal N}=2$ topological field theory
in $D=2$ \cite{Witten:1991zz}. The $\CN=2$ supersymmetry algebra
has two supersymmetry generators $Q_i, i=1,2$ that satisfy the
anticommutation relations \eqn{susyalg}{\{Q_{\alpha i}, Q_{\beta
j}\}=\delta_{ij} \gamma^\mu_{\alpha\beta}P_\mu.} In two
dimensions, the Lorentz group $SO(1,1)$ is generated by the
Lorentz boost $L.$ We diagonalize $L$ by going into the light-cone
frame $P_\pm=P_0\pm P_1,$ \eqalign{boostdiag}{[L,P_\pm]&=&\pm
P_\pm \cr \{L,Q_\pm\}&=&\pm\half Q_\pm.} The  commutation
relations of the $\CN=2$ supersymmetry algebra become
\eqalign{susydiag}{\{Q_{+i},Q_{+j}\}&=&\delta_{ij} P_+\cr
\{Q_{-i}, Q_{-j}\}&=&\delta_{ij} P_- \cr \{Q_{+i}, Q_{-j}\}&=&0.}
We let \eqn{brstcharge}{Q=Q_{+1}+iQ_{+2}+Q_{-1}\pm iQ_{-2}} with
either choice of sign.  It follows from \rf{susydiag} that $Q$ is
nilpotent \eqn{br}{Q^2=0,} so we  would like to consider $Q$ as a
BRST operator.

However $Q$ (\ref{brstcharge}) is not a scalar so this
construction would violate Lorentz invariance.  There is a way out
if the theory has left and right R-symmetries $R_+$ and $R_-.$
Under $R_+,$ the combination of supercharges $Q_{+1}\pm iQ_{+2}$
has charge $\pm1/2$ and $Q_{-1}\pm iQ_{-2}$ is neutral. For $R_-,$
the same is true with `left' and `right'  interchanged.

Hence, we can make $Q$ scalar if we modify the Lorentz generator
$L$ to be \eqn{lmod}{L'=L-\half R_+ \mp\half R_-.} At a more
fundamental level, this change in the Lorentz generator arises if
we replace the stress tensor $T_{\mu\nu}$ with \eqn{tmunu}{\tilde
T_{\mu\nu}= T_{\mu\nu}-\half(\partial_\mu J^+_\nu+\partial_\nu
J^+_\mu) \mp\half(\partial_\mu J^-_\nu+\partial_\nu J^-_\mu),}
where $J^+_\nu$ and $J^-_\mu$ are the left and right R-symmetry
currents. The substitution \rf{tmunu} is usually referred to as
`twisting' the stress tensor.

We give a new interpretation to the theory by taking $Q$ to be a
BRST operator. A state $\Psi$ is considered to be physical if it
is annihilated by $Q$ \eqn{qani}{Q\Psi=0.} Two states $\Psi$ and
$\Psi'$ are equivalent if \eqn{psiequi}{\Psi-\Psi'= Q\Phi,} for
some $\Phi.$ Similarly, we take the physical operators to commute
with the BRST charge \eqn{ocu}{[Q,{\cal O}]=0.} Two operators are
equivalent if they differ by an anticommutator of $Q,$
\eqn{eqvi}{{\cal O}'\sim {\cal O}+ \{Q,{\cal V}\},} for some
operator ${\cal V}.$

The theory with the stress tensor $\tilde T_{\mu\nu}$ and BRST
operator $Q$ is called a topological field theory. The basis for
the name is that one can use the supersymmetry  algebra to show
that the twisted stress tensor is BRST trivial
\eqn{trivial}{\tilde T_{\mu\nu}=\{Q,\Lambda_{\mu\nu}\}.} It
follows that in some sense the worldsheet metric is irrelevant.
The correlation function \eqn{cor}{\vev{\CO_1(x_1)
\CO_2(x_2)\dots\CO_n(x_n)}_\Sigma} of physical operators $\CO_i$
obeying $[Q,\CO_i]=0$ on a fixed Riemann surface $\Sigma$ is
independent of metric on $\Sigma.$ Indeed, varying the metric
$g_{\mu\nu}\rightarrow g_{\mu\nu}+\delta g_{\mu\nu},$ the
correlation function stays the same up to BRST trivial terms
\eqn{mvev}{\vev{\CO_1(x_1)\dots \CO_n(x_n)\int_\Sigma
\delta(\sqrt{g} g^{\mu\nu})\tilde T_{\mu\nu}}=\vev{\CO_1(x_1)\dots
\CO_n(x_n)\int_\Sigma\delta(\sqrt{g}
g^{\mu\nu})\{Q,\Lambda_{\mu\nu}\}}=0.}

More importantly for us, we can also construct a topological
string theory in which one obtains the correlation functions by
integrating \rf{cor} over the moduli of the Riemann surface
$\Sigma$ using $\Lambda_{\mu\nu}$ where the antighost $b_{\mu\nu}$
usually appears in the definition of the string measure.

For an $\CN=2$ supersymmetric  field theory in two dimensions with
anomaly-free left and right R-symmetries we get two topological
string theories, depending on the choice of sign in
(\ref{brstcharge}). We would like to consider the case that the
$\CN=2$ model is a sigma model with a target space being a complex
manifold $X.$ In this case, the two R-symmetries exist
classically, so classically we can construct the two topological
string theories, called the A-model and the B-model. Quantum
mechanically, however, there is an anomaly, and the B-model only
exists if $X$ is a Calabi-Yau manifold.

\subsection{Open String B-model on a Super-Twistor Space}

To define open strings in the B-model, one needs BRST invariant
boundary conditions. The simplest such conditions are Neumann
boundary conditions \cite{Witten:1992fb}. Putting in $N$ space
filling D5-branes gives $Gl(n,\Bbb C)$ (whose compact real form is
$U(N)$) gauge symmetry.  The zero modes of the D5-D5 strings give
a $(0,1)$ form gauge field $A=A_{\bar i}dz^{\bar i}$ in the target
space. The BRST operator acts as the $\bar\partial$ operator and
the string $\ast$ product is just the wedge product. Hence, $A$ is
subject to the gauge invariance \eqn{gaui}{\delta
A=Q\epsilon=\bar\partial \epsilon+[A,\epsilon],} and the string
field theory action reduces to the action of the holomorphic
Chern-Simons theory \cite{Witten:1992fb} \eqn{hcsc}{S={1\over
2}\int\Omega\wedge \Tr\left(A\wedge \bar\partial A+{2\over3}
A\wedge A\wedge A\right),} where $\Omega$ is the Calabi-Yau volume
form.

We would like to consider the open string B-model with target
space $\Bbb {CP}^3,$ but we cannot, since $\Bbb {CP}^3$ is not a
Calabi-Yau manifold and the B-model is well defined only on a
Calabi-Yau manifold. On a non-Calabi-Yau manifold, the R-symmetry
that we used to twist the stress tensor is anomalous. A way out is
to introduce spacetime supersymmetry. Instead of $\Bbb {CP}^3,$
which has homogeneous coordinates $Z^I, I=1,\dots, 4$ we consider
a supermanifold $\Bbb {CP}^{3|N}$ with bosonic and fermionic
coordinates \eqn{coordi}{Z^I,\quad \psi^A\qquad I=1,\dots, 4,
\quad A=1,\dots, N,} with identification of two sets of
coordinates that differ by a scaling \eqn{suca}{(Z^I,\psi^A)\cong
(tZ^I, t\psi^A) \qquad t\in \Bbb C^\ast.} The $\Bbb {CP}^{3|N}$ is
a Calabi-Yau supermanifold if and only if the number of fermionic
dimensions is $N=4.$ To see this, we construct the holomorphic
measure on $\Bbb {CP}^{3|4}.$ We start with the $(4|N)$ form on
$\Bbb C^{4|N}$ \eqn{ozero}{\Omega_0=dZ^1\dots dZ^4d\psi^1\dots
d\psi^N} and study its behavior under the rescaling symmetry
(\ref{suca}). For this, recall that $d\psi$ scales oppositely to
$\psi$ \eqn{suop}{(dZ^I, d\psi^A)\rightarrow (t dZ^I, t^{-1}
d\psi^A).} It follows, that $\Omega_0$ is $\Bbb C^\ast$ invariant
if and only if $N=4.$  In this case we can divide by the $C^\ast$
action and get a Calabi-Yau measure on $\Bbb {CP}^{3|4}$
\eqn{holf}{\Omega={1\over 4!}\epsilon_{IJKL}Z^I dZ^JdZ^KdZ^L
{1\over 4!}\epsilon_{ABCD} \psi^A\psi^B\psi^C\psi^D.}

The twistor space $\Bbb {CP}^3$ has a natural $Sl(4,\Bbb C)$ group
action that acts as $Z^A\rightarrow \Lambda^A{}_B Z^B$ on the
homogeneous coordinates of $\CP^3$. The real form $SU(2,2)$ of
$Sl(4,\Bbb C)$ is the conformal group of Minkowski space.
Similarly, the super-twistor space $\Bbb {CP}^{3|N}$ has a natural
$Sl(4|N,\Bbb C)$ symmetry. The real form $SU(2,2|N)$ of this is
the superconformal symmetry group with $N$ supersymmetries.

For $N=4$, the superconformal group $SU(2,2|4)$ is the symmetry
group of $\CN=4$ super-Yang-Mills theory.  In a sense, this is the
simplest non-abelian gauge theory in four dimensions. The $\CN=4$
superconformal symmetry uniquely determines the states and
interactions of the gauge theory. In particular, the beta function
of $\CN=4$ gauge theory vanishes.

Now we know a new reason for $\CN=4$ to be special. The
topological B-model on $\Bbb {CP}^{3|N}$ exists if and only if
$\CN=4.$  The B-model on $\Bbb {CP}^{3|4}$ has a $SU(2,2|4)$
symmetry coming from the geometric transformations of the twistor
space. This is related via the twistor transform to the $\CN=4$
superconformal symmetry.

In the topological B-model with space-filling branes on $\Bbb
{CP}^{3|4}$, the basic field is the holomorphic gauge field
$\CA=\CA_{\bar I} dZ^{\bar I}$, \eqn{abasic}{{\cal A}(Z,\bar Z,
\psi)= A(Z,\bar Z)+\psi^A \xi_{A}(Z,\bar Z)+{1\over
2!}\psi^A\psi^B \psi_{AB}(Z,\bar Z)+\dots+{1\over
4!}\epsilon_{ABCD}\psi^A\psi^B\psi^C\psi^D G(Z,\bar Z).} The
action is the same as \rf{hcsc}, except that the gauge field $\CA$
now depends on $\psi$ \eqn{actsu}{S=\half\int
\Omega\wedge\Tr\left(\CA \bar\partial \CA+
{2\over3}\CA\wedge\CA\wedge\CA\right),} and the holomorphic three
form is  (\ref{holf}).  The classical equations of motions
obtained from \rf{actsu} are
\eqn{cseq}{\bar\partial\CA+\CA\wedge\CA=0.} Linearizing the
equations of motions around the trivial solutions $\CA=0,$ they
tell us that \eqn{tela}{\bar\partial \Phi=0,} where $\Phi$ is any
of the components of $\CA.$ The gauge invariance reduces to
$\delta\Phi=\bar\partial\alpha.$ Hence for each component $\Phi,$
the field $\Phi$ defines an element of a cohomology group.

This action has the amazing property that its spectrum is the same
as that of $\CN=4$ super Yang-Mills theory in Minkowski space. To
see this, we need to use that the elements of  $(0,1)$ cohomology
groups of degree $2h-2$ are related by twistor transform to
helicity $h$ free fields on Minkowski space.

To figure out the degrees of various components $\CA$, notice that
the action must be invariant under the $\Bbb C^*$ action
$Z^I\rightarrow tZ^I.$ Since the holomorphic measure is also
invariant under the scaling, the only way that the action
\rf{actsu} is invariant is that the superfield $\CA$ is also
invariant, in other words, $\CA$ is of degree zero
\eqn{azero}{\CA\in H^{0,1}(\Bbb {CP}^{3|4},\CO(0)).} Looking back
at the expansion (\ref{abasic}) of the superfield, we identify the
components, via the twistor correspondence, with fields in
Minkowski space of definite helicity. $A$ is is of degree zero,
just like the superfield $\CA.$ Hence, it is related by twistor
transform to a field of helicity $+1.$ The field $G$ has degree
$-4$ to off-set the degree $4$ coming the four $\psi$, so it
corresponds to a field of helicity $-1.$ Continuing in this
fashion, we obtain the complete spectrum of $\CN=4$ supersymmetric
Yang-Mills theory. The twistor fields
$A,\xi_A,\phi_{AB},\xi_{ABC},G$  describe, via twistor transform,
particles of helicities $1,+\half,0,-\half,-1$ respectively.

The fields also have the correct representations under the $SU(4)$
R-symmetry group. This symmetry is realized in twistor space by
the natural geometric action on the fermionic coordinates
$\psi^A\rightarrow \Lambda^A{}_B \psi^B.$ Hence, $\psi^A$
transforms in the $\bf 4$ of the $SU(4)_R.$ The holomorphic gauge
superfield $\CA(Z,\psi)$ is invariant under the R-symmetry, hence
the representations of the components of $\CA$ must be conjugate
to the representations of the $\psi$ factors that they multiply in
\rf{abasic}. Hence, $A, \xi_{A}, \phi_{AB},\xi_{ABC}$ and $G$
transform in the $\bf 1, \bar 4, 6, 4, 1$ representation of
$SU(4)_R$ respectively\footnote{The construction of twistor string
has been generalized to theories with less supersymmetry or with
product gauge groups, by orbifolding the fermionic directions of
the super-twistor space \cite{Park:2004bw,Giombi:2004xv}.}

\subsection{ D-Instantons}

The action (\ref{actsu}) also describes some of the interactions
of $\CN=4$ super Yang-Mills, but not all. It cannot describe the
full interactions, because an extra $U(1)$ R-symmetry gets in the
way. The fermionic coordinates $\psi^A, A=1,\dots, 4$ have an
extra $U(1)_R$ besides the $SU(4)_R$ considered above. Indeed, the
full R-symmetry group in twistor space is
\eqn{rsym}{U(4)_R=SU(4)_R\times U(1)_R,} where we take the extra
$U(1)_R$,  which we call $S,$ to rotate the fermions by a common
phase \eqn{extrau}{S:\ Z^I\rightarrow Z^I,\quad \psi^A\rightarrow
e^{i\theta} \psi^A.} In the B-model, the extra $U(1)_R$ is
anomalous, since it does not leave fixed the holomorphic measure
$\Omega\sim d^3Z d\psi^1\dots d\psi^4.$ Under the $S$
transformation, the holomorphic  measure transforms as
$\Omega\rightarrow e^{-4i\theta}\Omega$, so it has charge $S=-4$,
hence the $B$-model action has $S=-4.$

However, as we have set things up so far, the anomaly of the
B-model is too trivial to agree with the anomaly of $\CN=4$
Yang-Mills theory. With the normalization \rf{extrau}, the $S$
charges of fields are given by their degrees. The $\CN=4$
Yang-Mills action is a sum of terms with $S=-4$ and $S=-8.$ For
illustration, consider the positive and negative helicity gluons
that have $S$-charge $0$ and $-4$ respectively. The kinetic term
and the $++-$ three-gluon vertex contribute to the $S=-4$ part of
the Yang-Mills action. The $--+$ and the $--++$ vertices
contribute to the $S=-8$ part.

The action of the open string B-model (\ref{actsu}) has $S=-4$
coming from the anomaly of $S$ of the holomorphic measure
$\Omega$. To get the $S=-8$ piece of the Yang-Mills action, we
need to enrich the B-model with nonperturbative instanton
contributions.

The instantons in question are Euclidean D1-branes wrapped on
holomorphic curves in $\Bbb {CP}^{3|4}$ on which open string can
end. The gauge theory amplitudes come from coupling of the open
strings to the D1-branes. The massless modes on the worldvolume of
a D-instanton are a $U(1)$ gauge field and the modes that describe
the motion of the instanton. In the following, we will study
mostly tree level amplitudes. These get contributions from genus
zero instantons on which the $U(1)$ line bundles have a fixed
degree $d=-1$. Hence the bundles do not have any discrete or
continuous moduli, so we will ignore the $U(1)$ gauge field from
now on. The modes describing the motion of the D-instanton make up
the moduli space ${\cal M}$ of holomorphic curves $C$ in the
twistor space. To construct scattering amplitudes we need to
integrate over $\CM.$

\section{Tree Level Amplitudes from Twistor String Theory}

\subsection{Basic Setup}
\label{basic}

Recall that the interactions of  the full gauge theory come from
Euclidean D1-brane instantons on which the open strings can end.
The open strings are described by the holomorphic gauge field
$\CA.$ To find the coupling of the open strings to the
D-instantons, let us consider the effective action of the D1-D5
and D5-D1 strings. Quantizing the zero modes of the D1-D5 strings
leads to a fermionic zero form field $\alpha^i$ living on the
worldvolume of the D-instanton. $\alpha^i$ transforms in the
fundamental representation of the $Gl(n,\Bbb C)$ gauge group
coming from the Chan-Paton factors. The D5-D1 strings are
described by a fermion $\beta_i$ transforming in the
antifundamental representation. The kinetic operator for the
topological strings is the BRST operator $Q,$ which acts as
$\bar\partial$ on the low energy modes. So the effective action of
the D1-D5 strings is \eqn{done}{S=\int_C \beta(\bar\partial+{\cal
A})\alpha,} where $C$ is the holomorphic curve wrapped by the
D-instanton. From this we read off the vertex operator for an open
string with wavefunction $\phi=\CA_{\bar I}dZ^{\bar I}$
\eqn{verterxop}{V_\phi=\int_C J\wedge \phi,} where $J_i{}^j=
\beta_i\alpha^j dz$ is  a holomorphic current made from the free
fermions $\alpha^j,\beta_i$. These currents generate a current
algebra on the worldvolume of the D-instanton.

To compute a scattering amplitude,  we evaluate the correlation
function \eqalign{conf}{\CA =\int d\CM \vev{
V_{\phi_1}V_{\phi_2}\dots V_{\phi_n}} =\int d\CM \left\langle
\int_C J_1\phi_1\dots \int_C J_n\phi_n \right\rangle.} We can
think of this as integrating out the fermions $\alpha,\beta$
living on the D-instanton. Hence, the generating function for
scattering amplitudes is simply the integral of Dirac operator
over moduli space of D-instantons \eqn{dii}{\int d\CM
\det(\bar\partial+ A).} Here,  $d{\cal M} $ is the holomorphic
measure on the moduli space of holomorphic curves of genus zero
and degree $d.$ In topological B-model, the action is holomorphic
function of the fields and all path integrals are contour
integral. Hence, the integral is actually over a
middle-dimensional Lagrangian cycle in the moduli space. This
integral is a higher dimensional generalization of the familiar
contour integral from complex analysis.

The correlator of the currents on D1-instanton\footnote{Here we
write the single trace contribution to the correlator that
reproduces the gauge theory scattering amplitude. As discussed in
section \ref{closed}, the multitrace contributions correspond to
gluon scattering processes with exchange of internal conformal
supergravity states.} \eqn{curc}{\langle J_1(z_1) J_2(z_2) \dots
J_n(z_n)\rangle={\Tr(T_1 T_2\dots T_n)dz_1 dz_2\dots dz_n\over
(z_1-z_2)(z_2-z_3)\dots (z_n-z_1)}+\,permutations} follows from
the free fermion correlator on a sphere
\eqn{fref}{\alpha^i(z)\beta_j(z')\sim {\delta^i{}_j\over z-z'}.}

\vskip .2in\noindent{\it Scattering Wavefunctions}

We would like to compute  the scattering amplitudes of plane waves
$\phi(x)=\exp\,(i \, p\cdot x)=\exp\,(i \, \pi^a \tilde\pi^{\dot
a} x_{a \dot a}).$ These are wavefunctions of external particles
with definite momentum ${p^{a \dot a}=\pi^a \tilde\pi^{\dot a}}.$
The twistor wavefunctions corresponding to plane waves are
\eqn{twav}{\phi(\lambda,\mu,\psi)= \bar\delta(\vev{\lambda,\pi})
\exp(i [{\tilde\pi},\mu])g(\psi),} where $g(\psi)$ encodes the
dependence on fermionic coordinates. For a positive helicity gluon
$g(\psi)=1$ and for a negative helicity gluon
$g(\psi)=\psi^1\psi^2\psi^3\psi^4.$ Here, we have introduced the
holomorphic delta function \eqn{hold}{\bar\delta(f)=\bar\partial
\bar f \delta^2(f),} which is a closed $(0,1)$ form. We normalize
it so that for any function $f(z)$, we have \eqn{delf}{\int dz
\,\bar\delta(z-a)f(z)=f(a).}

The idea of \rf{twav} is that the delta function
$\delta(\vev{\lambda,\pi})$ sets $\lambda^a$ equal to  $\pi^a.$
The Fourier transform of the exponential $\exp(i
[{\tilde\pi},\mu])$ back into Minkowski space gives another delta
function that sets $\tilde\lambda^{\dot a}$ equal to
$\tilde\pi^{\dot a}.$ The twistor string computation with these
wavefunctions gives directly momentum space scattering amplitudes.

Actually, the wavefunctions should be modified slightly so that
they are invariant under the scaling of the homogeneous
coordinates of $\CP^{3|4}.$ From the basic properties of delta
functions, it follows that $\bar\delta(\vev{\lambda,\pi})$ is
homogeneous of degree $-1$ in both $\lambda$ and $\pi.$ Hence, for
positive helicity gluons, the wavefunction is actually
\eqn{poh}{\phi^+(\lambda,\mu)=\bar\delta(\vev{\lambda,\pi})(\lambda/\pi)
\exp \big(i[{\tilde\pi},\mu](\pi/\lambda)\big).} Here,
$\lambda/\pi$ is a well defined holomorphic function, since
$\lambda$ is a multiple of $\pi$ on the support of the delta
function. The power of $(\lambda/\pi)$ was chosen, so that the
wavefunction is homogeneous of degree zero in overall scaling of
$\lambda,\mu,\psi.$  Under the scaling
\eqn{scals}{(\pi,\tilde\pi)\rightarrow (t\pi,t^{-1}\tilde\pi),}
the wavefunction is homogeneous of degree $-2$ as expected for a
positive helicity gluon \rf{scalia}. For negative helicity gluon,
the wavefunction is
\eqn{nev}{\phi^-(\lambda,\mu)=\bar\delta(\vev{\lambda,\pi})(\pi/\l)^3
\exp\big(i[\tilde\pi,\mu](\pi/\lambda)\big)
\psi^1\psi^2\psi^3\psi^4.} Under the scaling \rf{scals}, the
wavefunction is homogeneous of degree $+2$ as expected. For
wavefunctions of particles with helicity $h,$ there are similar
formulas with $2-2h$ factors of $\psi.$

\vskip .2in\noindent{\it MHV Amplitudes}

We saw in section \ref{twist} that MHV amplitudes, after Fourier
transform into twistor space, localize on genus zero degree one
curve $C$, that is, a linearly embedded copy of $\CP^1.$ Here we
will evaluate the degree one instanton contribution and confirm
that it gives the MHV amplitude.

Consider the moduli space of such curves. Each curve $C$ can be
described by the equations \eqn{dege}{\mu^{\dot a}=x^{a {\dot a}}
\lambda_a \qquad \psi^A=\theta^{Aa} \lambda_a,} where $\lambda^a$
are the homogeneous coordinates and $x^{a\dot a}$ and
$\theta^{Aa}$ are the moduli of $C$.  The holomorphic measure on
the moduli space is \eqn{meas}{d\CM=d^4x d^8\theta.} Hence, the
moduli space has $4$ bosonic and $8$ fermionic dimensions.

In terms of the homogeneous coordinate $\lambda^a$ the current
correlator \rf{curc} becomes
\eqn{cunu}{\vev{J_1(\pi_1)J_2(\pi_2)\dots
J_n(\pi_n)}={\prod_i\vev{\l_i,d\l_i}\over
\vev{\l_1,\l_2}\vev{\l_2,\l_3}\dots \vev{\l_n,\l_1}},} which we
found by setting $z_i=\l^2_i/\l^1_i$. We stripped away the color
factors and kept only the contribution of the term with
$1,2,\dots, n$ cyclic order. We multiply  this with the
wavefunctions
$\psi_i(\l,\mu,\psi)=\bar\delta(\vev{\l,\pi_i})\exp\left(i[\mu,\tilde\pi_i]\right)
g_i(\psi)$ and integrate over the positions $\l_i,\tl_i$ of the
vertex operators.  We perform the integral over the positions of
the vertex operators using the formula \eqn{homl}{\int_{\CP^1}
\vev{\l,d\l}\ \bar\delta(\vev{\l,\pi})f(\l)=f(\pi),} where $f(\l)$
is a homogeneous function of $\l^a$ of degree $-1.$ This is the
homogeneous version of definition of holomorphic delta function
\eqn{hdef}{\int_{\Bbb C} dz\ \bar\delta(z-b)f(z)=f(b).} Hence,
each wavefunction contributes a factor of \eqn{wco}{\int_C
\vev{\lambda,d\lambda}\phi_i=
\exp\left(i[\tilde\pi_i,\mu_i]\right)g_i(\psi_i),} where
$\mu_i^{\dot a}=x^{a\dot
a}\lambda_{ia},\psi^A_i=\theta^{aA}\lambda_{ia}$ and the delta
function sets $\lambda^a_i=\pi^a_i$ in the correlation function.
So the amplitude becomes \eqn{ale}{\CA={1\over \prod_k
\vev{\pi_k,\pi_{k+1}}} \int d^4x d^8\theta\ \exp\Big(i\sum_k
[\tilde\pi_k,\mu_k]\Big) \prod_k g_k(\psi_k).}

The fermionic part of the wavefunctions is $g_i=1$ for the
positive helicity gluons and
$g_i=\psi^1_i\psi^2_i\psi^3_i\psi^4_i$ for the negative helicity
gluons. Since we are integrating over eight fermionic moduli
$d^8\theta,$ we get nonzero contribution to amplitudes with
exactly two negative helicities $r^-,s^-.$ Setting
$\psi^A=\theta^{Aa}\pi_a,$ the integral over fermionic dimensions
of the moduli space gives the numerator of the MHV amplitude
\eqn{fermi}{\int d^8\theta \prod_{A=1}^4 \psi_r^A
\prod_{B=1}^4\psi_s^B= \vev{r,s}^4.} Setting $\mu^{\dot
a}_i=x^{a\dot a}\pi_{ia},\ i=1,\dots, n,$ the integral over
bosonic moduli gives the delta function of momentum conservation
\eqn{bosni}{\int d^4x \exp\left(ix_{a\dot
a}\sum_i\pi_i^a\tilde\pi^{\dot a}_i\right)=\delta^4(\sum_{i=1}^n
\pi_i^a\tilde\pi_i^{\dot a}).}

Collecting the various pieces, we get the familiar MHV amplitude
\eqn{mhva}{\CA(r^-,s^-)={\vev{r,s}^4 \over \prod_{i=1}^n
\vev{i,i+1}}\delta^4(\sum_{i=1}^n \pi_i\tilde\pi_i).}


\subsection{Higher Degree Instantons}
\label{highd}

\vskip .15in\noindent{\it Instanton Measure}

Here we will construct the measure on the moduli space of genus
zero degree $d$ curves. Such curves can be described as degree $d$
maps from an abstract $\CP^1$ with homogeneous coordinates $(u,v)$
\eqalign{didi}{Z^I&=&P^I(u,v) \cr \psi^A&=&\chi^A(u,v).} Here
$P^I,\chi^A$ are homogeneous polynomials of degree $d$ in $u,v.$
The space of homogeneous polynomials of degree $d$ is a linear
space of dimension $d+1,$ spanned by $u^d,u^{d-1}v,\dots, v^d.$
Picking a basis $b^\alpha(u,v),\alpha=1,\dots,d+1$, we write
\eqalign{expa}{P^I&=&\sum_\alpha P_\alpha^I\ b^\alpha\cr \psi^A&=&
\sum_\alpha \chi^A_\alpha\ b^\alpha.} A natural measure is
\eqn{namo}{d\CM_0=\prod_{\alpha=1}^{d+1}\prod_{A,I=1}^4
dP^I_\alpha\ d\chi^A_\alpha.} This measure is invariant under a
general $Gl(d+1,\Bbb C)$ transformation of the basis $b_\alpha.$
Since the number of bosonic and fermionic coordinates is the same,
the Jacobians cancel between fermionic and bosonic parts of the
measure. The description \rf{didi} is redundant, we need to divide
by the $\Bbb C^\ast$ action that rescales $P^I$ and $\chi^A$ by a
common factor. This reduces the space of curves from $\Bbb
C^{4d+4|4d+4}$ to $\Bbb{CP}^{4d+3|4d+4}.$ The curve $C$ also stays
invariant under an $Sl(2,\Bbb C)$ transformation on $(u,v)$ so the
actual moduli space of genus zero degree $d$ curves is
\eqn{modsp}{\CM=\Bbb{CP}^{4d+3|4d+4}/Sl(2,\Bbb C).}  As $d\CM_0$
is $Gl(2,\Bbb C)$ invariant, it descends to a holomorphic measure
\eqn{holm}{d\CM={d\CM_0\over Gl(2,\Bbb C)}.} on $\CM.$ Hence,
$\CM$ is a Calabi-Yau supermanifold of dimension $(4d|4d+4).$

We can now understand why amplitudes with different helicities
come from holomorphic curves of different degrees. Integrating
over the moduli space, the measure absorbs $4d+4$ fermion zero
modes. These come from the fermionic factors $g(\psi)$ in the
wavefunctions of the gluons \rf{twav}. A positive helicity gluon
does not contribute any zero modes while a negative helicity gluon
$g^-(\psi)=\psi^1\psi^2\psi^3\psi^4$ gives $4$ zero modes. Hence,
instantons of degree $d$ contribute to amplitudes with $d+1$
negative helicity gluons.

Alternatively, we can get this from counting the $S$ charge
anomaly. Wavefunctions of particles with different helicities
violate $S$ by different amount. The positive helicity gluons do
not violate $S$ while the negative helicity gluons violate $S$ by
$-4$ units. So, the amplitudes with $p$ positive helicity gluons
and $q$ negative helicity gluons violates the $S$ charge by $-4q$
units.

In the twistor string, there is a source of  violation of $S$ from
the instanton measure. Since the $S$ charge of $Z$ and $\psi$ is
$0$ and $1$ respectively, the charges of the coefficients
$P^I_\alpha, \chi^A_\alpha$ are $0,1.$ Hence, the differentials
$dP^I_\alpha, d\chi^A_\alpha$ have charges $0,-1$ and the $S$
charge of the $(4d|4d+4)$ dimensional measure $d\CM$ is $-4d-4.$

So an instanton can contribute to an amplitude with $q$ negative
helicity gluons if and only if \eqn{reld}{d=q-1.} This is the
familiar formula discussed at in subsection \ref{twist}. For $l$
loop amplitudes, this relation generalizes to $d=q-1+l.$

 \vskip
.15in\noindent{\it Evaluating the Instanton Contribution}

Here we consider the connected instanton contribution along the
lines of the calculation of the MHV amplitude. The amplitude is
\cite{Roiban:2004yf,Roiban:2004vt,Witten:2004cp}
\eqalign{wop}{\CA&=&\int d\CM_d \prod_i\int_C {\vev{u_i,du_i}
\over \prod_k \vev{u_k,u_{k+1}}}\bar\delta(\vev{\l(u_i),\pi_i})
\exp\left( i[\mu(u_i),\tilde\pi_i]\right) g_i(\psi_i).} Here
$d\CM_d$ is the measure on the moduli space of genus zero degree
$d$ curves. Next comes the correlator of currents on the
worldvolume of the D1-instanton and the wavefunctions in which we
use the parameterization $\lambda^a_i(u_i)=P^a(u_i), \mu^{\dot
a}(u_i)=P^{\dot a}(u_i).$

This is not really an integral. The integral over the $2d+2$
parameters $P^{\dot a}_\alpha$ gives $2d+2$ delta functions
because $P^{\dot a}$ appear only in the exponential
$\exp\left(\sum_i P(u_i)_{\dot a}\tilde\pi_i^{\dot a}\right)$.
Hence, we are left with an integral over $4d-(2d+2)+2n=2d+2n-2$
bosonic variables. Here the $2n$ integrals come from the
integration over the positions of the vertex operators. Now there
are $2n$ delta functions from the wavefunctions since each
holomorphic delta function is really a product of two real delta
functions $\bar\delta(z)=d\bar z\ \delta^2(z)$, and $2d+2$ delta
functions from the integral over the exponentials, which gives a
total of $2d+2n+2$. There are four more delta functions than
integration variables. The four extra delta functions impose
momentum conservation. Hence, the delta functions localize the
integral to a sum of contributions from a finite number of points
on the moduli space.

\vskip .15in\noindent{\it Parity Invariance}

In the helicity formalism, the parity symmetry of Yang-Mills
scattering amplitudes is apparent. The parity changes the signs of
the helicities of the gluons. The parity conjugate amplitude can
be obtained by simply exchanging $\lambda_i$'s with $\tl_i$'s.

To go to twistor space, one Fourier transforms with respect to
$\tl_i$, which breaks the symmetry between $\l$ and $\tl.$ Indeed,
the result \rf{wop} for the scattering amplitude treats $\l$ and
$\tl$ asymmetrically. An amplitude with $p$ positive helicities
and $q$ negative helicities has contribution from instantons of
degree $q-1$, while the parity conjugate amplitude with $q$ gluons
of positive helicity and $p$ gluons of negative helicity has
contribution from instantons of degree $p-1.$ To show that these
two are related by an exchange of $\l_i$ and $\tl_i$ requires some
amount of work. We refer the interested reader to the original
literature \cite{Roiban:2004yf,Roiban:2004vt,
Witten:2004cp,Berkovits:2004tx}.

\vskip .15in\noindent{\it Localization on the Moduli Space}

Recall that a tree level amplitude with $q$ negative helicity
gluons and arbitrary number of positive helicity gluons receives
contribution from instantons wrapping holomorphic curves of degree
$d=q-1.$ The degree $d$ instanton can consist of several disjoint
lower degree instantons whose degrees add up to $d.$ For
disconnected scattering amplitudes the instantons are connected by
open strings.
\begin{figure}
 \centering
 \includegraphics[height=1.6in]{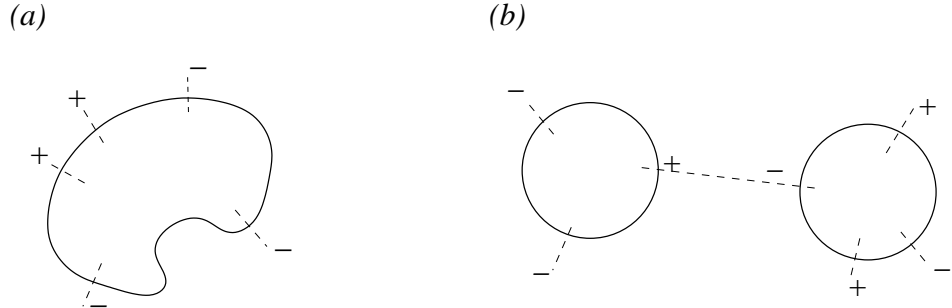}
 \caption{An amplitude with tree negative helicity gluons has contribution
          from two configurations: (a) Connected $d=2$ instanton. (b) Two disjoint
          $d=1$ instantons. The dashed line represents an open string
          connecting the instantons. }
 \label{coni}
\end{figure}
A priory, one expects that the amplitude receives contributions
from all possible instanton configurations with total degree
$q-1.$ So for example an amplitude with three negative helicity
gluons has contribution from a  connected $d=2$ instanton and a
contribution from two disjoint $d=1$ instantons, fig. \ref{coni}.

What one actually finds is that the connected and disconnected
instanton contributions reproduce the whole amplitude {\it
separately}. For example, in the case of amplitude with three
negative helicity gluons, it seems that there are two different
ways to compute the same amplitude. One can either evaluate it
from the connected $d=2$ instantons
\cite{Roiban:2004yf,Roiban:2004vt}, see fig. \ref{coni}(a).
Alternatively, the amplitude comes from evaluating the
contribution of the two disjoint $d=1$ instantons
\cite{Cachazo:2004kj}, fig. \ref{coni}(b).

\begin{figure}
 \centering
  \includegraphics[height=1.5in]{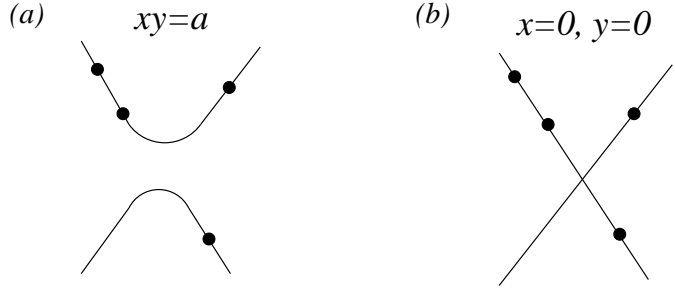}
 \caption{Localization of the connected instanton contribution to next to MHV amplitude;
  (a) the integral over the moduli space of connected degree two
  curves, localizes to an integral over the degenerate
  curves of (b), that is intersecting complex lines.
  In the figure, we draw the real section of the curves.}
 \label{locali}
\end{figure}

We can explain the equality of various instanton contributions
roughly as follows \cite{Gukov:2004ei}.  Consider the connected
contribution. The amplitude is expressed as a `contour' integral
over a middle-dimensional Lagrangian cycle in the moduli space of
degree two curves . The integrand comes from the correlation
function on the worldvolume of the D-instanton and from the
measure on the moduli space. It has poles in the region of the
moduli space, where the instanton degenerates to two intersecting
instantons of lower degrees $d_1+d_2=d$ , fig. \ref{locali}.
Picking a contour that encircles the pole, the integral localizes
to an integral over the moduli space $\CM'$ of the intersecting
lower degree curves.

Similarly, the  disconnected contribution has a pole when the two
ends of the propagator coincide. This comes from the pole of the
open string propagator \eqn{opro}{\bar\partial G={\bar
\delta}^3(Z'{}^I-Z^I) \delta^4(\psi'{}^A-\psi^A).} Hence, the
integral over disjoint instantons also localizes on the moduli
space of intersecting instantons. It can be shown that the
localized integrals coming from either connected or disconnected
instanton configurations agree \cite{Gukov:2004ei} which explains
why the connected and disconnected instanton calculations give the
entire scattering amplitude separately.

\vskip .2in \noindent{\it Towards MHV Diagrams}

Starting with a higher degree instanton contribution, successive
localization reduces the integral to the moduli space of
intersecting degree one curves. As we will review below, this
integral can be evaluated leading to a combinatorial prescription
for the scattering amplitudes \cite{Cachazo:2004kj}. Indeed,
degree one instantons give MHV amplitudes, so the localization of
the moduli integral leads to a diagrammatic construction based on
a suitable generalization of the MHV amplitudes.

\subsection{MHV Diagrams}

In this subsection, we start with a motivation of the MHV diagrams
construction of amplitudes from basic properties of twistor
correspondence. We then go on to discuss simple examples and
extensions to loop amplitudes. In the next subsection, we give a
heuristic derivation of the MHV rules from twistor string theory.

\begin{figure}
 \centering
 \includegraphics[height=1.5in]{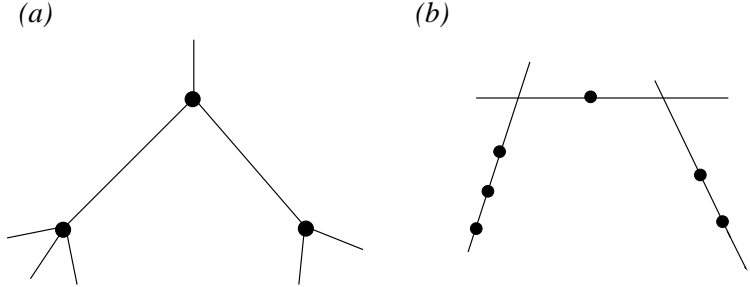}
 \caption{Two representations of a degree three MHV diagram. (a) In
          Minkowski space, the MHV vertices are represented by points.
          (b) In twistor space, each MHV vertex corresponds to a line.
          The three lines pairwise intersect.}
 \label{localize}
\end{figure}

Recall that MHV scattering amplitudes are supported on $\Bbb
{CP}^1$'s in twistor space. Each such $\Bbb {CP}^1$ can be
associated to a point $x^{a\dot a}$ in Minkowski space\footnote{We
are being slightly imprecise here. The space of $\Bbb {CP}^1$'s is
actually a copy of the complexified Minkowski space $\Bbb C^4.$
The Minkowski space $\Bbb R^{3|1}$ corresponds to $\Bbb {CP}^1$'s
that lie entirely in the 'null twistor space', defined by
vanishing of the pseudo-hermitian norm $Q(\lambda,\mu)=i(\lambda^a
\bar\mu_a-\bar\lambda^{\dot a}\mu_{\dot a}).$ Indeed, for a $\Bbb
{CP}^1$ corresponding to point a point in Minkowski space
$x^{a\dot a}$ is a hermitian matrix, hence it follows from
\rf{cpone} that $Q$ vanishes.} \eqn{cpone}{\mu_{\dot a}+x_{a\dot
a}\lambda^a=0.} So, in a sense, we can think of MHV amplitudes as
local interaction vertices \cite{Cachazo:2004kj}. To take this
analogy further, we can try to build more complicated amplitudes
from Feynman diagrams with vertices that are suitable off-shell
continuations of the MHV amplitudes, fig. \ref{localize}. MHV
amplitudes are functions of holomorphic spinors $\lambda_i$ only.
Hence, to use them as vertices in Feynman diagrams, we need to
define $\lambda$ for internal off-shell momenta $p^2\neq0.$

To motivate the off-shell continuation, notice that for on-shell
momentum $p^{a\dot a}=\lambda^a\tilde\lambda^{\dot a}$, we can
extract the holomorphic spinors $\lambda$ from the momentum $p$ by
picking arbitrary anti-holomorphic spinor $\eta^{\dot a}$ and
contracting it with $p^{a\dot a}.$ This gives $\lambda^a$ up to a
scalar factor \eqn{lax}{\lambda^a= {p^{a\dot a}\eta_{\dot a}\over
[\tilde\lambda, \eta]}.}  For off-shell momenta, this strategy
almost works except for the factor $[\tilde\lambda,\eta]$ in the
denominator which depends on the undefined spinor $\tilde\lambda$.
Fortunately, $[\tilde\lambda, \eta]$ scales out of Feynman
diagrams, so we take as our definition
\eqn{lambdaoffshell}{\lambda^a= p^{a\dot a}\eta_{\dot a}.} This is
clearly well-defined for off-shell momentum. We complete the
definition of the MHV rules, by taking $1/k^2$ for the propagator
connecting the MHV vertices.

Consider an MHV diagram with $v$ vertices. Each vertex gives two
negative helicity gluons. To make a connected tree level graph,
the vertices are connected with $v-1$ propagators. The propagators
absorb $v-1$ negative helicities, leaving $v+1$ negative helicity
external gluons. Hence, to find all MHV graphs contributing to a
given amplitude, draw all possible tree graphs of $v$ vertices and
$v-1$ links assigning opposite helicities to the two ends of
internal lines. The external gluons are distributed among the
vertices while preserving cyclic ordering. MHV graphs are those
for which each vertex has two negative helicity gluons emanating
from it.

For further work on MHV vertices construction of tree-level gluon
amplitudes, see
\cite{Bena:2004ry,Kosower:2004yz,Wu:2004fb,Zhu:2004kr,Birthwright:2005ak}.
MHV vertices have many generalizations; in particular, to
amplitudes with fermions and scalars
\cite{Georgiou:2004wu,Georgiou:2004by,Khoze:2004ba,Wu:2004jx,Su:2004ym},
with Higgses \cite{Dixon:2004za,Badger:2004ty} and with
electroweak vector-boson currents \cite{Bern:2004ba}. For an
attempt to generalize MHV vertices to gravity, see
\cite{Giombi:2004ix,Nair:2005iv,Abe:2005se}.

 \vskip .2in \noindent{\it Examples}

\begin{figure}
 \centering
  \includegraphics[height=1.8in]{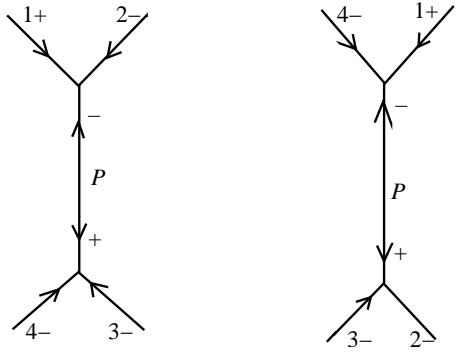}
 \caption{MHV diagrams contributing to the $+---$
          amplitude, which is expected to vanish.}
 \label{pmmm}
\end{figure}

Here we discuss concrete amplitudes to illustrate the method.
Consider first the $+---$ gluon amplitude. This amplitude vanishes
in Yang-Mills theory. It has contribution from two diagrams, see
fig. \ref{pmmm}.

The first of the two diagrams gives
\eqn{firstd}{{\vev{2,\lambda}^4 \over \vev{1,2} \vev{2,\lambda}
\vev{\lambda, 1}} {1 \over p^2} {\vev{3,4}^4 \over \vev{3,4}
\vev{4,\lambda} \vev{\lambda, 3}},} where we associate to the
internal momentum $p=p_1+p_2=-p_3-p_4$ the holomorphic spinor
\eqn{hof}{\lambda^a=p^{a\dot a}\eta_{\dot a}=(p_1+p_2)^{a {\dot
a}}\eta_{\dot a}.} The second diagram can be obtained from the
first by exchanging particles $2$ and $4$
\eqn{secd}{{\vev{\lambda', 4}^4\over\vev{1,
\lambda'}\vev{\lambda', 4}\vev{4, 1}}{1\over
p'^2}{\vev{2,3}^4\over\vev{2,3}\vev{3,\lambda'}\vev{\lambda',2}},}
where $\lambda'^a=(p_1+p_4)^{a\dot a}\eta_{\dot a}.$ Denoting
$\phi_i=\lambda_i^{\dot a}\eta_{\dot a},$ the first and second
diagrams  give respectively \eqn{firg}{-{\phi_1^3\over
\phi_2\phi_3\phi_4}{\vev{34}\over [21]}-{\phi^3_1\over
\phi_2\phi_3\phi_4}{\vev{32}\over[41]}.} The sum of these
contributions vanishes, because momentum conservation implies
$\vev{32}[21]+\vev{34}[41]=\sum_i \vev{3i}[i1]=0.$

It is easy to compute more complicated amplitudes. For example,
the $n$ gluon $---++\dots++$ amplitude is a sum of $2(n-3)$ MHV
diagrams, which can be obtained from fig. \ref{pmmm} by adding
additional $+$ helicities on the MHV vertices. The diagrams can be
evaluated to give \eqalign{nmhv}{ A&=&\sum_{i=3}^{n-1} {\vev{1
\lambda_{2,i}}^3 \over \vev{\lambda_{2,i} i+1} \vev{i+1 i+2} \dots
\vev{n1}}{1 \over q_{2i}^2} {\vev{23}^3 \over \vev{\lambda_{2,i}
2} \vev{34}\dots \vev{i\lambda_{2,i}}} \cr &+&\sum_{i=4}^n
{\vev{12}^3 \over \vev{2\lambda_{3,i}} \vev{\lambda_{3,i}i+1}
\dots \vev{n1}} {1\over q_{3i}^2} {\vev{\lambda_{3,i} 3}^3 \over
\vev{3,4} \dots \vev{i-1 i}\vev{i\lambda_{3,i}},}} where
$q_{ij}=p_i+p_{i+1}+\dots+p_j$ and the corresponding spinor
$\lambda_{i,j}^a$ is defined in the usual way
$\lambda^a_{i,j}=q_{ij}^{a\dot a}\ \eta_{\dot a}.$

\vskip .2in\noindent{\it Loop Amplitudes}

Similarly, one can compute loop amplitudes using MHV diagrams.
This has been carried out for the one loop MHV amplitude in
$\CN=4$ \cite{Brandhuber:2004yw} and $\CN=1$
\cite{Quigley:2004pw,Bedford:2004py} Yang-Mills theory, in
agreement with the known answers.

\begin{figure}
 \centering
 \includegraphics[height=1.2in]{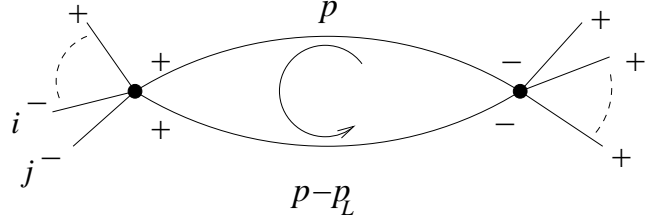}
 \caption{Schematic representation of a hypothetical twistor string computation of
          one-loop MHV amplitude. The picture shows a diagram in which the
           negative helicity gluons $i^-,j^-$ are on the same MHV vertex.}
 \label{mhv_loop}
\end{figure}

The expression for an MHV diagram  contributing to the one-loop
MHV amplitude is just what one would expect for a one-loop Feynman
diagram with MHV vertices, fig. \ref{mhv_loop}. There are two MHV
vertices, each coming with two negative helicity gluons. The
vertices are connected with two Feynman propagators that absorb
two negative helicities, leaving two negative helicity external
gluons \eqn{ampli}{\CA^{loop}=\sum_{{\cal D}, h} \int
{d^{4-2\epsilon} p\over (2\pi)^4} \CA_L(\lambda_k,
\lambda_p,\lambda_{p-p_L}){1\over
p^2(p-p_L)^2}\CA_R(\lambda_k,\lambda_p,\l_{p-p_L}).} The off-shell
spinors  entering the MHV amplitudes $\CA_L, \CA_R$ are determined
in terms of the momenta of the internal lines
\eqn{ofsh}{\l_p^a=p^{a\dot a}\eta_{\dot a},\qquad
\l_{p-p_L}^a=(p-p_L)^{a\dot a}\eta_{\dot a},} which is the same
prescription as for tree level MHV diagrams. The sum in \rf{ampli}
is over partitions ${\cal D}$ of the gluons among the two MHV
diagrams that preserve the cyclic order and over the helicities of
the internal particles\footnote{Similarly, the double-trace
contribution to one-loop MHV amplitudes comes from Feynman
diagrams with double-trace MHV vertices
\cite{Luo:2004ss,Luo:2004nw}.}.

\begin{figure}
 \centering
  \includegraphics[height=1.4in]{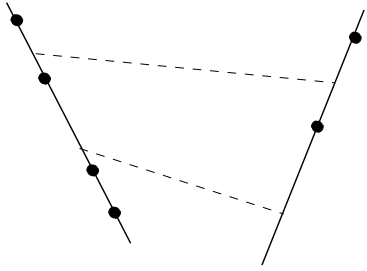}
   \caption{Twistor space structure of the one-loop MHV amplitude.
     The two MHV vertices are represented by lines. In a hypothetical
     twistor string computation of the amplitude, the lines are
     connected by two twistor propagators to make a loop.}
 \label{loop_twistor}
\end{figure}

This calculation makes the twistor structure of one-loop MHV
amplitudes manifest. The two MHV vertices are supported on  lines
in twistor space, so the amplitude is a sum of contributions, each
of which is supported on a disjoint union two lines. In a
hypothetical twistor string theory computation of the amplitude,
these two lines are connected by open string propagators, see fig.
\ref{loop_twistor}. This pictures agrees with studies of the
twistor structure using differential equations
\cite{Cachazo:2004zb}, after taking into account the holomorphic
anomaly of the differential equations
\cite{Cachazo:2004by,Bena:2004xu}.

Finally, we make a few remarks about the nonsupersymmetric
one-loop MHV amplitudes. The $\CN=0$ MHV amplitudes are sums of
cut-constructible terms and  rational terms. The cut-constructible
terms are correctly reproduced from MHV diagrams
\cite{Bedford:2004nh}.  The rational terms are single valued
functions of the spinors, hence they are free of cuts in four
dimensions. Their twistor structure suggests that they receive
contribution from diagrams in which, alongside with MHV vertices,
there are new one-loop vertices coming from one-loop all-plus
helicity amplitudes \cite{Cachazo:2004zb}. However, a suitable
off-shell continuation of the one-loop all-plus amplitude has not
been found yet. There has been recent progress in computing the
rational part of some one-loop QCD amplitudes using a
generalization \cite{Bern:2005hs} of the tree level recursion
relations reviewed in section \ref{perturb}.

\subsection{Heuristic Derivation of MHV Diagrams from Twistor String Theory}

\begin{figure}
 \centering
  \includegraphics[height=1.2in]{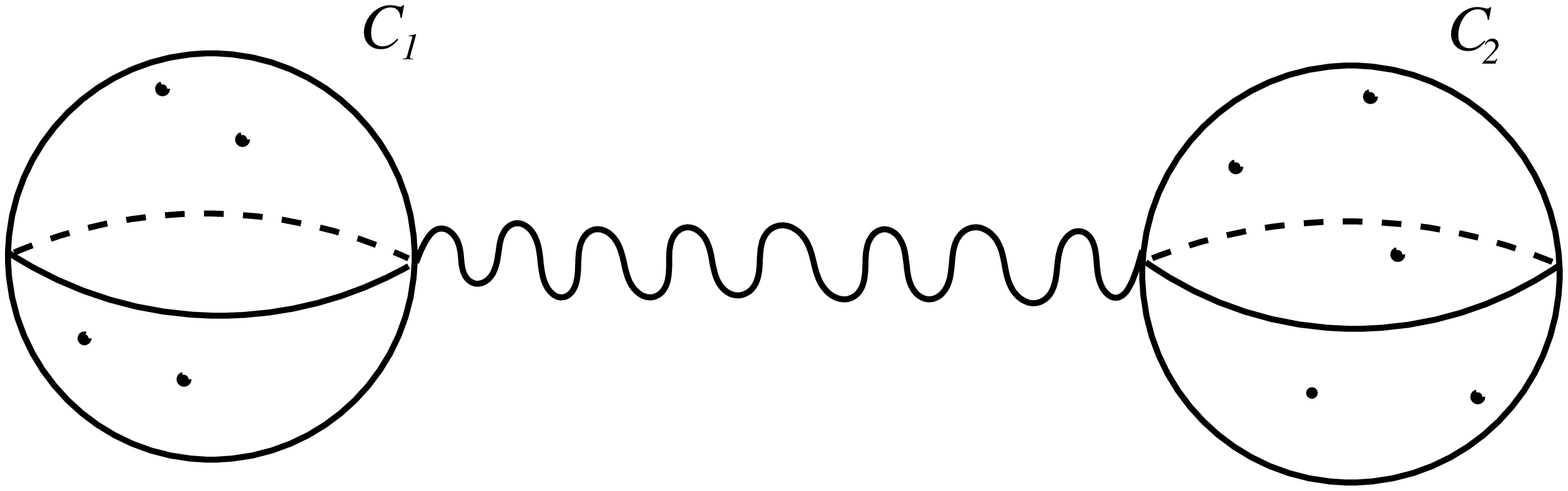}
 \caption{Twistor string contribution to an amplitude with three negative
 helicity external gluons.
 Two disconnected degree one instantons are connected by an open string.}
 \label{thret}
\end{figure}

Here, we will make an analysis of the disconnected twistor
diagrams that contribute to tree level amplitudes\footnote{For an
attempt to derive MHV rules from $\CN=4$ superspace constraints,
see \cite{Abe:2004ep}.}. We will evaluate the twistor string
amplitude corresponding the twistor contribution of fig.
\ref{thret} and show how it leads to the MHV diagrammatic rules of
the last subsection.

The physical field of the open string B-model is a $(0,1)$-form
$\CA$ with kinetic operator  $\bar\partial$ coming from the
Chern-Simons action. The twistor propagator for $\CA$ is a
$(0,2)$-form on $\Bbb {CP}^3\times \Bbb{CP}^3$ that is a
$(0,1)$-form on each copy of $\Bbb{CP}^3.$ The propagator obeys
the equation \eqn{opror}{\bar\partial G={\bar
\delta}^3(Z_2^I-Z_1^I) \delta^4(\psi_2^A-\psi_1^A).} Here,
$\bar\delta(z)=d{\bar z} \delta(z)\delta({\bar z})$ is the
holomorphic delta function $(0,1)$-form.

In an axial gauge, the twistor propagator becomes
\eqn{tiwp}{G=\bar\delta(\l_2^2-\l_1^2)\bar\delta(\mu_2^{\dot
1}-\mu_1^{\dot 1}){1\over \mu_2^{\dot 2}-\mu_1^{\dot
2}}\prod_{A=1}^4(\psi_2^A-\psi_1^A),} where we set
$\l_1^1=\l_2^1=1.$

For simplicity, we evaluate the contribution from two degree-one
instantons $C_1$ and $C_2$ connected by twistor propagator. This
configuration contributes to amplitudes with three negative
helicity gluons. The instantons $C_i, i=1,2$ are described by the
equations \eqn{insto}{\mu^{\dot a}_k=x_i^{a {\dot a}}
\lambda_{ka}, \qquad \psi_k^A=\theta^{Aa}_i\lambda_{ka}\qquad
i=1,2,\ k=1,\dots, n.} Here, $x^{a {\dot a}}_i$ and
$\theta^{Aa}_i$ are the bosonic and fermionic moduli of $C_i$.

With our choice of gauge, the twistor propagator is supported on
points such that $\lambda^a_1=\lambda^a_2.$ Since $\mu^{\dot
a}_2-\mu^{\dot a}_1=y^{a\dot a}\lambda_a,$ where $y^{a\dot
a}=x^{a\dot a}_2-x^{a\dot a}_1,$ the condition $\mu^{\dot
1}_2-\mu^{\dot 1}_1=0$ implies $\lambda^a=y^{a\dot 1}.$ Hence, the
bosonic part of the propagator gives $1/(\mu^{\dot 2}_2-\mu^{\dot
2}_1)=1/y^2.$

The  correlators of the gluon vertex operators on $C_1$ and $C_2$
and the integral over $\theta^{Aa}_i$ give two MHV amplitudes
$\CA_L$ and $\CA_R$ as explained in the $d=1$ computation. So we
are left with the integral \eqn{dtwoi}{\int d^4x_1 d^4x_2 \CA_L
{1\over (x_2-x_1)^2} \CA_R \prod_{i \in L} \exp(i x_1 \cdot p_i)
\prod_{j \in R} \exp(i x_2 \cdot p_j),} where the integral is over
a suitably chosen  $4\times4$ real dimensional `contour' in the
moduli space $\Bbb C^4\times \Bbb C^4$ of two degree one curves.
We rewrite the exponential as \eqn{expo}{\exp(i y \cdot P)
\prod_{j \in L,R} \exp(i x \cdot p_i),} where $x\equiv x_1$ and
$P=\sum_{i \in R} p_i$ is momentum of the off-shell line
connecting the two vertices. The integral \eqn{dem}{\int d^4x
\prod_{i \in L,R} \exp(i x \cdot p_i)=(2\pi)^4\delta^4(\sum_i
p_i)} gives the delta function of momentum conservation. We are
left with \eqn{abem}{A= \int d^4 y{1 \over y^2} \exp(i y \cdot P)
\CA_L\CA_R.}  The integrand has a pole at $y^2=0,$ which is the
condition for the curves $C_1$ and $C_2$ to intersect. The space
$y^2=0$ is the familiar conifold. It is a cone over $\Bbb
{CP}^1\times \Bbb {CP}^1$ so we parameterize it as \eqn{cop}{y^{a
{\dot a}}=t \lambda^a \tilde\lambda^{\dot a}.} Here $\lambda^a\in
{\cal O}(1,0),\, \tilde\lambda^{\dot a} \in {\cal O}(0,1),$ hence
$t\in {\cal O}(-1,-1)$ so that \rf{cop} is well-defined. We choose
a contour that picks the residue at $y^2=0.$ The residue is the
volume form on the conifold \eqn{refa}{ {\rm Res} \, {d^4 y \over
y^2} = t dt \vev{\lambda, d\lambda} [\tilde\lambda,
d\tilde\lambda].} Taking the residue, the integral becomes
\eqn{aboc}{ I= \int t dt\vev{\lambda, d\lambda} [\tilde\lambda,
d\tilde\lambda] \exp(i t P_{a\dot a}\lambda^a \tilde\lambda^{\dot
a})\CA_L \CA_R,} where the MHV vertices depend on the holomorphic
spinor $\lambda$ only. We pick the contour $t\in(-\infty,\infty)$
and ${\tilde\lambda=\bar\lambda},$ that is, we integrate over the
real light-cone. For ${t\in(0,\pm\infty)}$ we regulate the
integral with the prescription ${P=(p^0\pm i\epsilon,\vec{p})},$
so \eqn{princ}{{\int_{-\infty}^\infty t dt \exp(i tP_{a\dot
a}\lambda^a \tilde\lambda^{\dot a})=-{2\over (P_{a\dot a}\lambda^a
\tilde\lambda^{\dot a})^2}}.} Hence we have \eqn{next}{{I= \int
\vev{\lambda, d\lambda} [\tilde\lambda, d\tilde\lambda] {1 \over
(P \lambda \tilde\lambda)^2} \CA_L\CA_R(\lambda)}.} To reduce the
integral \rf{next} to a sum over MHV diagrams, we use the identity
\eqn{rein}{{ {[\tilde\lambda,d\tilde\lambda] \over (P \lambda
\tilde\lambda )^2}= -{1\over P\lambda \eta} \, \bar\partial \left(
{[\tilde\lambda, \eta] \over P \lambda \tilde\l } \right)},} where
$\eta^{\dot a}$ is an arbitrary positive helicity spinor to write
the integral as \eqn{ipre}{{I=\int \vev{\lambda ,d\lambda} \,
{\CA_L \CA_R \over (P\lambda \eta)} \bar \partial \left(
{[\tilde\lambda, \eta] \over (P \lambda \tilde\lambda)} \right)}.}
Now we can integrate by parts. The $\bar\partial$ operator acting
on the holomorphic function on the left gives zero except for
contributions coming from  poles of the holomorphic function,
$\bar\partial\left(1/z\right)=\bar\delta(z).$ These evaluate to a
sum over residues \eqn{byp}{{I=\sum {\rm Res} \, \left(
{\CA_L\CA_R \over P\lambda \eta} \right) {[\tilde\lambda, \eta]
\over P \lambda \tilde\lambda}}.} The residues of
${1/(P\lambda\eta)}$ are at \eqn{pol}{\lambda^a=P^{a\dot
a}\eta_{\dot a}.} Substituting this back into \rf{byp},
$P\lambda\tilde\lambda$ evaluates to $P^2 [\tilde\l,\eta],$ so we
have \eqn{mhve}{{I={1 \over P^2} \CA_L\CA_R(\lambda=P\eta)}.} But
this is precisely the contribution from an MHV diagram. Summing
over all cyclicly ordered partitions of the gluons among the two
instantons gives the sum over MHV diagrams contributing to the
scattering amplitude.

\begin{figure}
 \centering
  \includegraphics[height=1.6in]{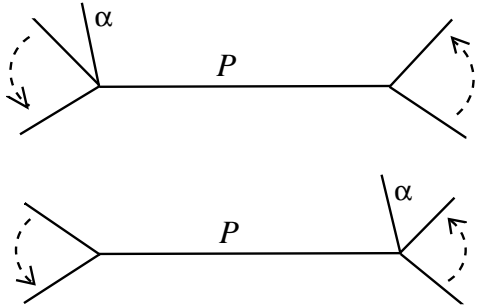}
 \caption{The graphs contributing to the pole at $\lambda=\lambda_\alpha.$
          The reversed order of $\alpha$ and the internal line in the two graphs,
          changes the sign of the residue of the pole.}
 \label{cancellation}
\end{figure}

There are additional additional poles in \rf{byp} that come from
the MHV vertices $\CA_L\CA_R$ \eqn{apo}{{{1 \over
\prod_{\alpha=1}^4 \vev{\lambda_\alpha, \lambda}}},} where
$\alpha$ runs over the four gluons  adjacent to the twistor line.
The poles are located at  ${\lambda=\lambda_\alpha},$ which is the
condition of the twistor line to meet the gluon vertex operator.
Consider the two diagrams, fig. \ref{cancellation} in which the
function $\CA_L\CA_R$ has a pole at $\lambda=\lambda_\alpha.$ The
graphs differ by whether the gluon $\alpha$ is on the left vertex
just after the propagator or on the right vertex just before the
propagaor. The reversed order of ${\lambda}$ and
${\lambda_\alpha}$ in the two diagrams changes the sign of the
residue. The rest of the residue \rf{byp} stays the same after
taking ${\lambda=\lambda_\alpha}.$ The off-shell momenta of the
two diagrams differ by ${\delta P=\lambda_\alpha
\tilde\lambda_\alpha},$ so the diagrams have the same value of the
denominators ${(P\lambda_\alpha
\tilde\lambda_\alpha)(P\lambda_\alpha \eta)}.$ Hence, all poles at
${\lambda=\lambda_\alpha}$ get cancelled among pairs of diagrams.

This derivation clearly generalizes to several disconnected degree
one instantons that contribute to a general tree level amplitude.
An amplitude with $d+1$ negative helicity gluons gets
contributions from diagrams with $d$ disconnected degree one
instantons. The evaluation of the twistor contributions leads to
MHV diagrams with $d$ MHV vertices.

Let us remark that the integral \rf{next} could be taken as the
starting point in the study of MHV diagrams.  Since \rf{princ} is
clearly Lorentz invariant,\footnote{The Lorentz invariance
requires some elaboration, because the choice of contour
$\bar\lambda=\tilde\lambda,$ breaks the complexified Lorentz group
$Sl(2,\Bbb C)\times Sl(2,\Bbb C)$ to the diagonal $Sl(2,\Bbb C),$
the real Minkowski group. It can be argued from the holomorphic
properties of the integral \rf{next}, that it is invariant under
the full $Sl(2,\Bbb C)\times Sl(2,\Bbb C)$ \cite{Cachazo:2004kj}.}
the MHV diagram construction must be Lorentz invariant as well.
Although separate MHV diagrams depend on the auxiliary spinor
$\eta,$ the sum of all diagrams contributing to a given amplitude
is independent of the auxiliary spinor $\eta^{\dot a}$.

\vskip .2in \noindent{\it Loops in Twistor Space?}

We have just seen that the disconnected instanton contribution
leads to tree level MHV diagrams. However, the MHV diagram
construction seems to work for loop amplitudes as well, as
discussed in previous subsection. Hence, one would like to
generalize the twistor string derivation to higher genus instanton
configurations, which contribute to loop amplitudes in Yang-Mills
theory. For example, the one-loop MHV amplitude should come from a
configuration of two degree one instantons connected by two
twistor propagators to make a loop, fig. \ref{loop_twistor}. An
attempt to evaluate this contribution runs into difficulties: the
two twistor propagators are both inserted at the same point
$\lambda^a=y^{a\dot a}\eta_{\dot a}$ on the D-instanton
worldvolume making the answer ill-defined. Some of these
difficulties are presumably related to the closed string sector of
the twistor string theory, that we will now review.


\section{Closed Strings}
\label{closed}

The closed strings of the topological B-model on supertwistor
space are related by twistor transform to $\CN=4$ conformal
supergravity \cite{Berkovits:2004jj}. The conformal group is the
group of linear transformations of the twistor space, so the
twistor string is manifestly conformally invariant. Hence it
necessarily leads to a conformal theory of gravity.

\subsection{Closed String Spectrum}

Let us see how the closed strings are related to the conformal
supergravity fields. The most obvious closed string field is the
deformation of complex structure of $\CP'^{3|4},$ the $'$ means
that we throw away the set $\lambda^a=0.$ In this and the
following section, we parameterize $\CP^{3|4}$ with homogeneous
coordinates $Z^I,\ I=1,\dots,8.$ Recall that the complex structure
is conventionally defined in terms of the tensor field
$j^A=J^A{}_B dZ^B$ obeying $J^2=-1.$ The indices $A,B$ can be both
holomorphic or antiholomorphic. In local holomorphic coordinates
$J^I{}_J=i$ and $J^{\bar I}{}_{\bar J}=-i$. The first order
perturbations of the complex structure are described by a field
$J^I{}_{\bar J}$ and its complex conjugate $J^{\bar I}{}_J$. From
$J^I{}_{\bar J}$ we form the vector valued $(0,1)$ form
$j^I=J^I{}_{\bar J} dZ^{\bar J}$ with equations of motion
\eqn{ejej}{\bar\partial j^I=0} that express the integrability
condition on the deformed complex structure. $j^I$ is volume
preserving $\partial_I J^I_{\bar J}=0$, since the holomorphic
volume $\Omega$ is part of the definition of the
B-model\footnote{This extra condition is not understood from
B-model perpective \cite{Berkovits:2004jj}. One can guess it from
analogous condition in the Berkovits's open twistor string.},
$j^I$ is subject to the gauge symmetry $j^I\rightarrow
j^I+\epsilon \bar\partial \kappa^I$, where $\kappa^I$ is a volume
preserving vector field.

According to twistor transform \cite{Penrose:1976jq}, volume
preserving deformations of complex structure of twistor space are
related to anti-selfdual perturbations of the spacetime.
Anti-selfdual perturbations correspond to positive helicity
conformal supergravitons. The $\CN=4$ positive helicity
supermultiplet contains fields going from the helicity $+2$
graviton to a complex scalar $\bar C.$

The negative helicity graviton is part of a separate $\CN=4$
superfield. It comes from an RR two form field
\eqn{bfield}{b=B_{\bar I J}\ d\bar Z^{\bar I}\wedge dZ^J} that
couples to the D1-branes of the B-model via \eqn{bcop}{\int_C b,}
where $C$ is the worldvolume of the D1-brane. The equations of
motion of $b$ are \eqn{eomg}{\bar\partial b=0} and $b$ is subject
to the gauge invariance $b\rightarrow b+\bar\partial \lambda.$ In
order to relate $b$ to the fields of the Berkovits's open twistor
string that we discuss in next section, one needs to assume that
$b$ is also invariant under the gauge transformation $B_{\bar
IJ}\rightarrow B_{\bar IJ}+\partial_J\ \chi_{\bar I}.$

\subsection{Conformal Supergravity}

Conformal supergravity in four dimensions has action
\eqn{weyla}{S\sim\int d^4x \sqrt{-g}W_{abcd}W^{abcd},} where $W$
is the Weyl tensor. This theory is generally considered
unphysical. Expanding the action around flat space
$g_{\mu\nu}=\eta_{\mu\nu}+h_{\mu\nu}$ leads to a fourth order
kinetic operator $S\sim\int d^4x\, h
\partial^4h$ for the fluctuations of the metric, and thus to
lack of unitarity.

\begin{figure}
 \centering
 \includegraphics[height=1.1in]{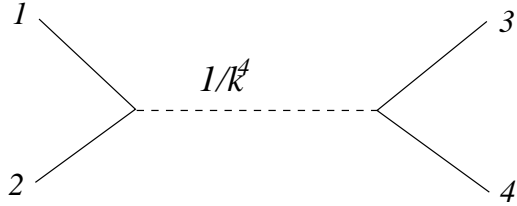}
 \caption{A double trace  $ \Tr T_1T_2\Tr T_3T_4$ contribution to
          tree level four gluon scattering amplitude coming from
          exchange of conformal supergravity particle,
          which is represented by a dashed line.}
 \label{graviton}
\end{figure}

We can see a sign of the supergravity already in the tree level
MHV amplitude calculation of section \ref{basic}. There we found
that the single trace terms agree with the tree level MHV
amplitude in gauge theory. We remarked that the current algebra
correlators give additional multi-trace contributions. These come
from an exchange of an internal conformal supergravity state,
which is a singlet under the gauge group\footnote{These
open-closed string interactions can be used to study deformations
of $\CN=4$ gauge theory by turning on closed string background
field \cite{Kulaxizi:2004pa,Chiou:2005jn}.}. For example, the four
gluon MHV amplitude has a contribution $\Tr\ T_1T_2 \Tr\ T_3T_4$
coming from an exchange of supergravity state in the
$12\rightarrow 34$ channel, fig. \ref{graviton}. In twistor string
theory, this comes from the double trace contribution of the
current algebra on the worldvolume of the D-instanton
\eqn{twiam}{\int_{\CM} d\CM \left\langle V_1 V_2\right\rangle
\left\langle V_3 V_4\right\rangle.}

At tree level, it is possible to recover the pure gauge theory
scattering amplitudes by keeping the single-trace terms. However,
at the loop level, the diagrams that include conformal
supergravity particles can generate single-trace interactions.
Hence the presence of conformal supergravity coming from the
closed strings puts an obstruction to computation of Yang-Mills
loop amplitudes in the present formulation of twistor string
theory.

In twistor string theory, the conformal supergravitons have the
same coupling as gauge bosons, so it is not possible to remove the
conformal supergravity states by going to weak coupling. Since,
Yang-Mills theory is consistent without conformal supergravity, it
is likely that there is a version of the twistor string theory
that does not contain the conformal supergravity states.

\section{Berkovits's Open Twistor String}
\label{berko}

Here we will describe the open string version of the twistor
string \cite{Berkovits:2004hg}. In this string theory, both
Yang-Mills and conformal supergravity states come from open string
vertex operators.

\subsection{The Spectrum}

The action of the open string theory is \eqn{opac}{S=\int
d^2z\left( Y_I\bar\nabla_{\bar z} Z^I+\bar Y_I \nabla_z\bar
Z^I+S_C \right).} For Euclidean signature of the worldsheet,
$Z^I=(\l^a,\tl^{\dot a},\phi^A),\ a,{\dot a}=1,2,\ A=1,\dots,4,\
I=1,\dots,8,$ are homogeneous coordinates on $\CP^{3|4}$ and $\bar
Z^I$ are their complex conjugates. $Y_I$ and $\bar Y_I$ are
conjugates to $Z^I$ and $\bar Z^I.$ Notice that $Z^I,I=5,\dots, 8$
were denoted as $\phi^A,A=1,\dots, 4$ in previous sections. Before
twisting, $Z$ and $\bar Z$ have conformal weight zero and $Y$ and
$\bar Y$ have conformal weight one. The covariant derivatives are
\eqn{covd}{\nabla_z=\partial_z-A_z\qquad \nabla_{\bar
z}=\bar\partial_{\bar z}-A_{\bar z},} where $A$ is a worldsheet
gauge field that gauges the $Gl(1,\Bbb C)$ symmetry
$Z^I\rightarrow tZ^I,\ Y_I\rightarrow t^{-1} Y_I.$ $S_C$ is the
action of a  current algebra with central charge $+28$ which
cancels $-26$ of the conformal ghosts and $-2$ of the $Gl(1,\Bbb
C)$ ghosts.  The open string boundary conditions are
\eqn{bondc}{Z^I=\bar Z^I,\quad Y_I=\bar Y_I\quad j_r=\bar j_r,}
where $j_r,\ r=1,\dots \dim G$ are the currents of the current
algebra. On the boundary, $Z$ and $Y$ are real and the $Gl(1,\Bbb
C)$ gauge group is broken to the  group $Gl(1,\Bbb R)$ of real
scalings of $Z,Y.$

The physical open string vertex operators are described by
dimension one fields that are neutral under $Gl(1)$ and primary
with respect to Virasoro and $Gl(1)$ generators
\eqn{genp}{T=Y_I\partial Z^I+T_C,\qquad J=Y_IZ^I.} The fields
corresponding to Yang-Mills states are
\eqn{vym}{V_\phi=j_r\phi^r(Z)} where $\phi^r(Z)$ is a dimension
zero $Gl(1,\Bbb R)$ neutral function of $Z.$ That is, $\phi$ is
any function on $\RP^{3|4}.$ $V_\phi$ has clearly dimension one.
$\phi$ is related by twistor transform to gauge fields on
spacetime with signature $++--.$

The vertex operators describing the conformal supergravity are
\eqn{vertg}{V_f=Y_If^I(Z),\qquad V_g=\partial Z^I g_I(Z).} These
have dimension one, since $Y_I$ and $\partial Z^I$ have dimension
one. The $Gl(1)$ invariance requires that $f^I$ has $Gl(1)$ charge
$+1$ and $g_I$ has $Gl(1)$ charge $-1.$ The vertex operators are
primary if \eqn{pric}{\partial_I f^I=0,\qquad Z^Ig_I=0.}  We
identify two vertex operators that differ by null states
\eqn{din}{\delta V_f=J_{-1}\Lambda=Y_IZ^I\Lambda,\qquad \delta
V_\phi=T_{-1}\chi=\partial Z^I\partial_I\chi.} Hence, $f^I$ and
$g_I$ are subject to the gauge invariance \eqn{fgga}{\delta
f^I=Z^I\Lambda,\qquad \delta g_I=\partial_I \chi.}

Since $f^I$ has $Gl(1)$ charge $+1,$ we can construct $Gl(1)$
neutral the vector field \eqn{ups}{\Upsilon=f^I{\partial\over
\partial Z^I}.}  $\Upsilon$ descends to
a vector field on on the real twistor space $\Bbb R^{3|4}$ thanks
to the gauge invariance $\delta f^I=Z^I\Lambda$ that kills the
vertical part of the vector field along the $Gl(1)$ orbits
$Z^I\rightarrow tZ^I.$ The primary condition $\partial_I f^I=0$
implies that $\Upsilon$ preserves the volume measure $\Omega\sim Z
d^7Z$ on $\RP^{3|4}.$ Hence $\Upsilon$ is a volume preserving
vector field on $\RP^{3|4}$. Similarly, we can summarize the
conditions on $g$ by considering the $1$ form
\eqn{oneg}{\Theta=g_I dZ^I.} The constraint $g_I Z^I=0$ means that
$\Theta$ annihilates the vertical vector field
$Z^I\partial/\partial Z^I,$ so it descends to a one form on
$\RP^{3|4}.$ The gauge invariance $\delta g_I=\partial_I \chi$
means that $\Theta$ is actually an abelian gauge field on
$\RP^{3|4}.$

\vskip .2in\noindent{\it Comparison with B-model}

Recall that that B-model is defined on $\CP'^{3|4}$. The open
strings correspond to gauge fields in Minkowski space and the
closed strings correspond to conformal supergravity. On the other
hand in the open twistor string both gauge theory and conformal
supergravity states come from the open string vertex operators.
The boundary of the worldsheet (and hence the vertex operators)
lives in $\RP'^{3|4}$. Hence the twistor fields are related by
twistor transform to fields on spacetime with signature $++--.$

The gauge field is described in B-model by a $(0,1)$ form $\CA$
that is an element of $H^1(\CP'^{3|4},\CO)$. This has equations of
motion $\bar\partial \CA=0$ and gauge invariance
$\delta\CA=\bar\partial\epsilon,$ where $\epsilon$ is a function
on $\CP'^{3|4}$. In open string, the gauge field comes from a
function $\phi$ on $\RP'^{3|4}.$ If $\phi$ is real-analytic, we
can extend it to a complex neighborhood of $\RP'^{3|4}$ in
$\CP'^{3|4}$. Then the relation between the two fields is
\cite{Atiyah:1979iu,Berkovits:2004jj,Witten:2003nn}
\eqn{reat}{\CA=\phi\ \bar\partial(\theta\left({\rm Im}\ z\right))=
{i\over 2} \phi\ \delta\left({\rm Im}\ z\right) d\bar z,} where
$z=\lambda^2/\lambda^1$ and $\theta(x)=1$ for $x\geq0$ and $0$ for
$x<0.$

The B-model closed string field giving deformation of complex
structure $j^I=J^I{}_{\bar J} dZ^J$ is related to the open string
volume preserving vector field $\Upsilon=f^I\ {\partial/\partial
Z^I}$ as $j^I= f^I\ \bar\partial(\theta\left({\rm Im}\ z
\right)).$ Similarly, the RR-two form $b=B_{I\bar J}\ dZ^I\wedge
dZ^{\bar J}$ gets related to the abelian gauge field $\Theta=g_I\
dZ^I$ of the open string by $b=\Theta\
\bar\partial(\theta\left({\rm Im}\ z \right)).$

Hence, we get the open twistor string wavefunction by considering
$\lambda^a,\mu^{\dot a}$ real and  by replacing holomorphic delta
functions $\bar\delta(\vev{\l,\pi})$ with real delta functions
\eqn{bwa}{\phi(\lambda,\mu,\psi)=\delta(\vev{\lambda,\pi})
\exp\left(i[\tilde\pi,\mu]\right) g(\psi).}

\subsection{ Tree Level Yang-Mills Amplitudes}

A tree level $n$ gluon scattering amplitude has contribution from
worldsheet $D$ of disk topology. The gluon vertex operators are
inserted along the boundary of the worldsheet. Taking the disk to
be the upper half-plane ${\rm Im}\ z\geq 0$, we insert the vertex
operators at $z_i,\ {\rm Im}\ z_i=0.$ Hence, the scattering
amplitude is \eqn{scatr}{\CA=\sum_d\int d\CM  \vev{\int dz_1
V_\phi(z_1) \dots \int dz_n V_\phi(z_n)},} where the sum is over
$U(1)$ worldsheet instantons and $d\CM$ is the measure.

In two dimensions, the instanton number of a $Gl(1)$ gauge bundle
is the degree of the line bundle. Recall that the  bundle $\CO(d)$
of degree $d$ homogeneous functions has degree $d$. Hence, on a
worldsheet with instanton number $d$, $Z^I$'s are sections of
$\CO(d).$ But this is just the parametric description of an
algebraic curve of degree $d$ discussed in section \ref{highd}.
While in B-model we summed over D-instantons, in the open twistor
string we are summing over  worldsheet instantons. Both
description lead to the same curves in twistor space. The only
difference is that for B-model we consider holomorphic curves,
while here we are interested in real algebraic curves.

The discussion of the real case is entirely analogous to the
holomorphic case.  Each $Z^I$ has $d+1$ real zero modes that are
local coordinates on the moduli space
$\CM=\RP^{4d+3|4d+4}/Sl(2,\Bbb R).$ The measure is just the
holomorphic measure \rf{holm} restricted to real curves. The
moduli space of degree $d$ instantons has $4d+4$ fermionic
dimensions. Since negative helicity gluon gives $4$ zero modes and
positive helicity gluon gives no zero modes, a degree $d$
instanton contributes to amplitudes with $d+1$ negative helicity
gluons. Parameterizing the disk using $z,\ {\rm Im} z\geq 0,$ the
amplitude is the real version of \rf{wop}
\eqalign{rwop}{\CA&=&\int d\CM_d \prod_i\int_D {dz_i \over \prod_k
(z_k-z_{k+1})} \delta(\vev{\l(z_i),\pi_i}) \exp\left(
i[\mu(z_i),\tilde\pi_i]\right) g_i(\psi_i).} In
\cite{Berkovits:2004tx}, a cubic open string field theory was
constructed for the Berkovits's twistor string theory. Since the
twistor string field theory gives the correct cubic
super-Yang-Mills vertices, it provides further support that
\rf{rwop} correctly computes tree-level Yang-Mills amplitudes.

\section{Recent Results in Perturbative Yang-Mills}
\label{perturb}

In this part of the lecture we shift gears and concentrate on new
techniques for the calculation of scattering amplitudes in gauge
theory. We will discuss two main results: BCFW recursion relations
\cite{Britto:2004ap, Britto:2005fq} for tree amplitudes of gluons
and quadruple cuts of ${\cal N}=4$ one-loop amplitudes of gluons
\cite{Britto:2004nc}.

\subsection{BCFW Recursion Relations}

We have seen how tree-level amplitudes of gluons can be computed
in a simple and systematic manner by using MHV diagrams. However,
from the study of infrared divergencies of one-loop ${\cal N}=4$
amplitudes of gluons, surprisingly simple and compact forms for
many tree amplitudes were found in \cite{Bern:2004ky,
Roiban:2004ix}. These miraculously simple formulas were given an
explanation when a set of recursion relations for amplitudes of
gluons was conjectured in \cite{Britto:2004ap}. The
Britto-Cachazo-Feng-Witten (BCFW) recursion relations were later
proven and extended in \cite{Britto:2005fq}. Here we review the
BCFW proof of the general set of recursion relations. The reason
we choose to spend more time in the proof than in recursion
relation itself is that the proof is constructive and the same
method can and has been applied to many other problems from field
theory to perhaps string theory.

Consider a tree-level amplitude $\CA (1,2,\ldots ,n-1,n)$ of $n$
cyclically ordered gluons, with any specified helicities. Denote
the momentum of the $i^{th}$ gluon by $p_i$ and the corresponding
spinors by $\lambda_i$ and $\tilde\lambda_i$. Thus, $p_i^{a\dot a}
=\lambda_i^a\tilde\lambda_i^{\dot a}$, as usual in these lectures.

In what follows, we single out two of the gluons for special
treatment.  Using the cyclic symmetry, without any loss of
generality, we can take these to be the gluons $k$ and $n$. We
introduce a complex variable $z$, and let
\eqalign{deftwo}{ p_{k}(z) & = \lambda_{k}(\tilde\lambda_{k} -
z\tilde\lambda_n), \cr p_n(z) & = (\lambda_n +
z\lambda_{k})\tilde\lambda_n. }  We leave the momenta of the other
gluons unchanged, so $p_s(z)=p_s $ for $s\not= k,n$. In effect, we
have made the transformation
\eqn{thetrans}{\tilde\lambda_k\to\tilde\lambda_k-z\tilde\lambda_n,\quad
~~\lambda_n\to\lambda_n+z\lambda_k,}
with $\lambda_k$ and $\tilde\lambda_n$ fixed. Note that $p_k(z)$
and $p_n(z)$ are on-shell for all $z$, and $p_k(z)+p_n(z)$ is
independent of $z$. As a result, we can define the following
function of a complex variable $z$,
\eqn{defz}{\CA(z) = \CA(p_1,\ldots ,p_{k-1}, p_{k}(z),
p_{k+1},\ldots ,p_{n-1}, p_n(z)). }
The right hand side is a physical, on-shell amplitude for all $z$.
Momentum is conserved and all momenta are on-shell.

For any $z\not=0$, the deformation \rf{deftwo} does not make sense
for real momenta in Minkowski space, as it does not respect the
Minkowski space reality condition $\tilde\lambda = \pm
\bar\lambda$. However, \rf{deftwo} makes perfect sense for complex
momenta or  (if $z$ is real) for real momenta in signature
$++-\,-$.   In any case, we think of $\CA(z)$ as an auxiliary
function. In the end, all answers are given in terms of spinor
inner products and are valid for any signature.

Here we assume that the helicities $(h_k,h_n)$ are $(-,+)$. The
proof can be extended to helicities $(+,+)$, or $(-,-)$ but we
refer the reader to \cite{Britto:2005fq}.

We claim three facts about $\CA(z)$: $(1)$ It is a rational
function. $(2)$ It only has simple poles. $(3)$ It vanishes for
$z\to \infty$.

These three properties of $\CA(z)$ imply that it can be written as
follows
\eqn{ratio}{\CA(z) = \sum_{p\in \{ {\rm poles}\} }
\frac{c_p}{z-z_p},}
where $c_p$ is the residue at a given pole and the sum is over the
whole set of poles. It turns out that, as we will see below, $c_p$
is proportional to the product of two physical amplitudes with
fewer gluons than $\CA(z)$. Therefore, \rf{ratio} provides a
recursion relation for amplitudes of gluons.

Let us prove the three statements. $(1)$ This is easy. Note that
the original tree-level amplitude is a rational function of spinor
products. Since the $z$ dependence enters only via the shift
$\tilde\lambda_k \to \tilde\lambda_k - z\tilde\lambda_n$ and
$\lambda_n \to \lambda_n + z \lambda_k$, $\CA(z)$ is clearly
rational in $z$.

$(2)$ By definition, $\CA(z)$ is constructed out of Feynman
diagrams. The only singularities $\CA(z)$ can have come from
propagators. Recall that $\CA(z)$ is color-ordered. This means
that all propagators are of the form $1/P^2_{ij}$ where
$P_{ij}=p_i+\ldots + p_j$. Clearly, $P_{ij}$ is $z$ independent if
both $k,n\in \{ i,\ldots, j\}$ or if $k,n\not\in \{ i,\ldots,
j\}$. By momentum conservation it is enough to consider
propagators for which $n\in \{ i,\ldots, j\}$ and $k \not\in \{
i,\ldots, j\}$. Since the shift of $p_n$ is by a null vector, one
has
\eqn{porpi}{ P_{ij}^2(z)  = P^2_{ij}(0) - z
\langle\lambda_k|P_{ij}|\tilde \lambda_n],}
where for any spinors $\lambda,\tilde\lambda$ and vector $p$, we
define $\langle \lambda|p|\tilde\lambda]=- p_{a\dot
a}\lambda^a\tilde\lambda^{\dot a}$. Hence, the propagator
$1/P_{ij}(z)^2$ has only a single, simple pole, which is located
at $z_{ij}= P_{ij}^2/\langle\lambda_k|P_{ij}|\tilde\lambda_n]$.

$(3)$ Recall that any Feynman diagram contributing to the
amplitude $\CA(z)$ is linear in the polarization vectors
$\epsilon_{a\dot a}$ of the external gluons. Polarization vectors
of gluons of negative and positive helicity and momentum $p_{a\dot
a}= \lambda_a \tilde\lambda_{\dot a}$ can be written respectively
as follows (see section \ref{spinors} ),
\eqn{poli}{\epsilon_{a\dot a}^{-} = {\lambda_a\tilde\mu_{\dot
a}\over [\tilde\lambda , \tilde\mu ] }, \qquad \epsilon_{a\dot
a}^{+} = {\mu_a \tilde\lambda_{\dot a}\over \vev{\mu, \lambda}},}
where $\mu$ and $\tilde\mu$ are fixed reference spinors.

Only the polarization vectors of gluons $k$ and $n$ can depend on
$z$. Consider the $k^{th}$ gluon first. Recall that $\lambda_k$
does not depend on $z$ and $\tilde\lambda_k(z)$ is linear in $z$.
Since $h_k=-1$, it follows from \rf{poli} that $\epsilon^{-}_k$
goes as $1/z$ as $z\to \infty$. A similar argument leads to
$\epsilon^{+}_n \sim 1/z$ as $z\to \infty$.

The remaining pieces in a Feynman diagram are the propagators and
vertices. It is clear that the vanishing of $\CA(z)$ as $z\to
\infty$ can only be spoiled by the momenta from the cubic
vertices, since the quartic vertices have no momentum factors and
the propagators are either constant or vanish for $z\to\infty$.

Let us now construct the most dangerous class of graphs and show
that they vanish precisely as $1/z$. The $z$ dependence in a tree
diagram ``flows" from the $k^{th}$ gluon to the $n^{th}$ gluon
along a unique path of propagators. Each such propagator
contributes a factor of $1/z$.  If there are $r$ such propagators,
the number of cubic vertices through which the $z$-dependent
momentum flows is at most $r+1$. (If all vertices are cubic, then
 starting from the $k^{th}$ gluon, we
find a cubic vertex and then a propagator, and so on. The final
cubic vertex is then joined to the $n^{th}$ gluon.) So the
vertices and propagators give a factor that grows for large $z$ at
most linearly in $z$.

As the product of polarization vectors vanishes as $1/z^2$, it
follows  that for this helicity configuration, $\CA(z)$ vanishes
as $1/z$ for $z\to \infty$.

Now we can rewrite \rf{ratio} more precisely as follows
\eqn{collo}{\CA(z)=\sum_{i,j}{c_{ij}\over z-z_{ij}},}
where $c_{ij}$ is the residue of $\CA(z)$ at the pole $z=z_{ij}$.
From the above discussion, the sum over $i$ and $j$ runs over all
pairs such that $n$ is in the range from $i $ to $j$ while $k$ is
not. At this point it is clear the smallest number of poles is
achieved when $k$ and $n$ are adjacent, i.e., $k=n-1$. This is the
choice we make in the examples below.

Finally, we have to compute the residues $c_{ij}$. To get a pole
at $P_{ij}^2(z)=0$, a tree diagram must contain a propagator that
divides it into a ``left'' part containing all external gluons not
in the range from $i$ to $j$, and a ``right'' part containing all
external gluons that are in that range. The internal line
connecting the two parts of the diagram has momentum $P_{ij}(z)$,
and we need to sum over the helicity $h=\pm$ at, say, the left of
this line. (The helicity at the other end is opposite.)  The
contribution of such diagrams near $z=z_{ij}$ is $\sum_h
\CA_L^h(z)\CA_R^{-h}(z)/P_{ij}(z)^2$, where $\CA_L^h(z)$ and
$\CA_R^{-h}(z)$ are the amplitudes on the left and the right with
indicated helicities. Since the denominator $P_{ij}(z)^2$ is
linear in $z$, to obtain the function $c_{ij}/(z-z_{ij})$ that
appears in \rf{collo}, we must simply set $z$ equal to $z_{ij}$ in
the numerator. When we do this, the internal line becomes
on-shell, and the numerator becomes a product
$\CA_L^h(z_{ij})\CA_R^{-h}(z_{ij})$ of physical, on-shell
scattering amplitudes. More precisely we have,
\eqn{preci}{\CA_L^h(z_{ij}) = \CA(p_{j+1},\ldots
,p_k(z_{ij}),\ldots , p_{i-1}, P^{h}_{ij}(z_{ij})), \quad
\CA_R^{-h}(z_{ij}) = \CA(-P^{-h}_{ij}(z_{ij}),p_i,\ldots,
p_n(z_{ij}),\ldots , p_j). }

The formula \rf{collo} for the function $\CA(z)$ therefore becomes
\eqn{gollo}{\CA(z)=\sum_{i,j}\sum_h{\CA_L^h(z_{ij})\CA_R^{-h}(z_{ij})\over
P_{ij}(z)^2}.} To get the physical scattering amplitude
$\CA(1,2,\ldots, n-1,n)$, we set $z$ to zero in the denominator
without touching the numerator. Hence,
\eqn{wollo}{\CA(1,2,\ldots,
n-1,n)=\sum_{i,j}\sum_h{\CA_L^h(z_{ij})\CA_R^{-h}(z_{ij})\over
P_{ij}^2}.}
This is the BCFW recursion relation
\cite{Britto:2004ap,Britto:2005fq}.

\subsubsection{Examples}

Let us illustrate some of the compact formulas one can obtain
using the recursion relations \rf{wollo}.

Consider two of the six-gluon next-to-MHV amplitudes, for example,
amplitudes with three minus and three plus helicity gluons: $
\CA(1^-,2^-,3^-,4^+,5^+,6^+)$ and $\CA(1^+,2^-,3^+,4^-,5^+,6^-)$.
As mentioned above, the recursion relations \rf{wollo} have the
smallest number of terms when $k$ and $n$ are chosen to be
adjacent gluons. In the first example we choose to shift $p_3$ and
$p_4$, while in the second we shift $p_2$ and $p_3$. The results
are the following:
\eqn{lepta}{ \CA(1^-,2^-,3^-,4^+,5^+,6^-) = {1\over \gb{ 5 | 3+4 |
2}}\left( {\gb{ 1 | 2+3 | 4}^3 \over [2~3][3~4]\vev{5~6}\vev{6~1}
t_2^{[3]}} + {\gb{ 3 | 4+5 | 6}^3 \over
[6~1][1~2]\vev{3~4}\vev{4~5} t_3^{[3]} }\right).}
\eqalign{alos}{ \CA(1^+,2^-,3^+,4^-,5^+,6^-) &=& \frac{ [1~3]^4
\vev{4~6}^4 }{ [1~2] [2~3] \vev{4~5} \vev{5~6} t_{1}^{[3]}
\gb{6|1+2|3} \gb{4|2+3|1}} \\ &+& \frac{ \vev{2~6}^4 [3~5]^4 }{
\vev{6~1}\vev{1~2} [3~4] [4~5] t_{3}^{[3]} \gb{6|4+5|3}
\gb{2|3+4|5}} \\ &+& \frac{ [1~5]^4 \vev{2~4}^4 }{
\vev{2~3}\vev{3~4} [5~6] [6~1] t_{2}^{[3]} \gb{4|2+3|1}
\gb{2|3+4|5}},}
where $t_i^{[r]} = p_i + \ldots + p_{i+r-1}$.

It is interesting to observe that while \rf{lepta} and \rf{alos}
are simpler than the amplitudes computed by Berends, Giele,
Mangano, Parke, Xu \cite{Berends:1987me, Mangano:1987xk,
Berends:1989hf, Mangano:1990by}; the former possess spurious
poles, like $\gb{ 5 | 3+4 | 2}$, while the latter only have
physical poles. One can use the recursion relations to find
further simple formulas for tree-level gluon amplitudes
\cite{Britto:2005dg}.

Also note that the two-term form \rf{lepta} was obtained in
\cite{Roiban:2004ix} as a collinear limit of a very compact form
of the seven-gluon amplitude, which was originally obtained from
the infrared behavior of a one-loop ${\cal N}=4$ amplitude
\cite{Bern:2004ky}.

Let us also mention that many generalizations of the BCFW
recursion relations have been made, in particular, to include
amplitudes with fermions and scalars \cite{Luo:2005rx,Luo:2005my}
and to gravity amplitudes \cite{Bedford:2005yy, Cachazo:2005ca}.
The recursion relations have also been generalized to amplitudes
with massive particles \cite{Badger:2005zh}, and to some one-loop
amplitudes in QCD \cite{Bern:2005hs}.

\subsection{One-Loop ${\cal N}=4$ Amplitudes of Gluons and Quadruple Cuts}

Supersymmetric amplitudes of gluons are very special. The main
reason is that these amplitudes are four-dimensional
cut-constructible. This means that a complete knowledge of their
branch cuts and discontinuities, when the dimensional
regularization parameter is taken to zero, is enough to determine
the full amplitude. This is not true for non-supersymmetric
amplitudes. As an example consider the one-loop
$\CA(1^+,2^+,\ldots , n^+)$ and $\CA(1^-,2^+,\ldots , n^+)$
amplitudes. One can prove that both series of amplitudes are
single valued functions of the kinematical invariants. This is
enough to conclude that they vanish in any supersymmetric
theory\footnote{This can also be derived using supersymmetric Ward
identities. For a nice review see \cite{Dixon:1996wi}.}. In
contrast, in non-supersymmetric gauge theories, they are
interesting rational functions. These two series of amplitudes
were shown to be reproduced by a generalization of the BCFW
recursion relations in \cite{Bern:2005hs}.

Here we concentrate on ${\cal N}=4$ one-loop amplitudes. The
reason these amplitudes are special within the class of
supersymmetric amplitudes is that they can be expressed in terms
of known scalar box integrals with coefficients that are rational
functions of the spinor products.

In this part of the lectures, we explain a new technique that
allows the computation of any given scalar box coefficient as the
product of four tree-level amplitudes. Now recall that either by
using MHV diagrams or the BCFW recursion relations\footnote{We
also need the corresponding generalizations to include fermions
and scalars \cite{Georgiou:2004by, Wu:2004fb, Wu:2004jx,
Georgiou:2004wu}.}, any tree-level amplitude can be easily
computed. This implies that the new technique solves the problem
of computing one-loop amplitudes of gluons in ${\cal N}=4$ super
Yang-Mills.

\subsubsection{Review of The Unitarity-Based Method}

One of the most successful methods in the calculation of one-loop
amplitudes of gluons is the unitarity-based method
\cite{Bern:1994zx, Bern:1994cg}. This method was used to calculate
all MHV amplitudes \cite{Bern:1994zx} and all six-gluon
next-to-MHV amplitudes \cite{Bern:1994cg} more than a decade ago.
We review the basic idea of the method focusing on the points that
prepare the ground for the quadruple cut method.

The unitarity-based method can be described as a three-step
procedure: (1) Consider a given amplitude and use
Passarino-Veltman or other reduction techniques \cite{passarino}
to find a set of basic integrals. In supersymmetric amplitudes of
gluons, this means that any tensor Feynman integrals that enters
in a Feynman diagram calculation can be reduced to a set of scalar
integrals, that is Feynman integrals in a scalar field theory with
a massless particle running in the loop, with rational
coefficients. In particular, for ${\cal N}=4$ super Yang-Mills,
only scalar box integrals appear.

Scalar box integrals are defined as follows,
\eqn{ifou}{I_{(K_1,K_2,K_3,K_4)} = \int d^4\ell { 1\over
(\ell^2+i\epsilon) ((\ell-K_1)^2 + i\epsilon )((\ell -K_1-K_2)^2 +
i\epsilon )((\ell+K_4)^2 + i\epsilon )}.}
This is really a function of only three momenta $K_1, K_2, K_3$,
for $K_4=-K_1-K_2-K_3$ by momentum conservation. This integral is
UV finite but it has IR divergencies when at least one $K_i$ is
null, i.e., $K_i^2=0$. This implies that a regularization
procedure, like dimensional regularization, is required. The
structure of the IR singular terms is well understood
\cite{Catani:1998bh}. We do not discuss it here because it is not
relevant for the quadruple cut technique.

In a given amplitude, $K_i$ is the sum of consecutive momenta of
external gluons. We discuss this in more detail below.

(2) Consider a unitarity cut in a given channel, say the
$s-$channel. Recall that this is defined by summing over all
Feynman diagrams that contain two propagators whose momenta differ
by $s$ and by cutting those two propagators. Cutting a propagator
$1/(P^2+i\epsilon)$ means removing the principal part, i.e.,
replacing the propagator by $\delta^{(+)} (P^2)$. When this is
done, the internal particles go on-shell and the sum over Feynman
diagrams produces two tree-level amplitudes while the integration
over the internal momenta becomes an integration over the Lorentz
invariant phase space of two null vectors; this is known as a cut
integral. As an example, consider the cut in the
$P_{ij}^2$-channel, see figure \rf{uni}, the cut integral is given
by \cite{cuts}
\eqn{cutii}{C = \int d\mu ~ \CA^{\rm tree} (\ell_1,i,...,j,
\ell_2) ~ \CA^{\rm tree}(-\ell_2,j+1,..., i-1,-\ell_1),}
where $d\mu$ is the Lorentz invariant phase space measure for
$(\ell_1,\ell_2)$. The measure is explicitly given by
\eqn{mme}{ d\mu = d^4\ell_1 d^4\ell_2 \delta^{(+)} (\ell_1^2)
\delta^{(+)} (\ell_2^2) \delta^{(4)}(\ell_1 + \ell_2 - P_{ij}),}
with $P_{ij}$ denoting the sum of the momenta of gluons from $i$
to $j$.

\begin{figure}
\centering
\includegraphics[height=1.7in]{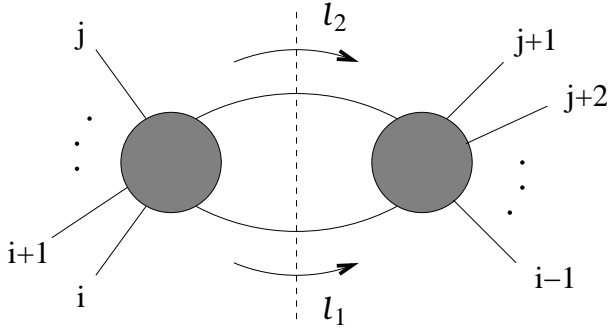}
\caption{Unitarity cut in the $P_{ij}^2-$channel.
The blobs represent tree-level amplitudes in which the propagator
lines are interpreted as external on-shell particles.} \label{uni}
\end{figure}

(3) Use reduction techniques to write the integrand of \rf{cutii}
as a sum of terms that contain a constant coefficient times two
propagators. Once this is achieved, it is easy to construct a
function of scalar box integrals with given coefficients that has
such a cut. Then repeat this for all other cuts, remembering that
a given scalar box integral has cuts in several different
channels. This means that one should not just add the functions
obtained from the study of each channel. Instead one has to
combine them while avoiding to overcount. Once a function with all
the correct discontinuities has been constructed, this must be the
final answer for the amplitude. The reason is that supersymmetric
amplitudes are four-dimensional cut-constructible, as mentioned
above.

Using this technique, all MHV amplitudes and the six-gluon NMHV
amplitudes were computed more than ten years ago. More recently,
the seven-gluon NMHV amplitude with all minus helicity gluons
adjacent was computed by using a combination of this method and
the holomorphic anomaly of unitarity cuts \cite{Cachazo:2004by} in
\cite{Cachazo:2004dr, Britto:2004nj}. The same result as well as
all other helicity configurations for the seven-gluon amplitude
were obtained by the unitarity-based method in \cite{Bern:2004ky}.

At this point it is important to mention that the integrand in the
cut integral \rf{cutii} is complicated because in general there
are many scalar box integrals sharing the same branch cut. The
reduction techniques, though systematic, can lead to very large
expressions for the scalar box coefficients \cite{Bern:2004ky}.
These large expressions can be shown to be equivalent to simple
formulas obtained as educated guesses \cite{Bern:2004ky}. This is
a hint that there must be a more direct method for computing such
coefficients.

A related difficulty comes from the fact that a given scalar box
integral has many different branch cuts. This means that after its
coefficient has been computed from a given cut, one still has to
disentangle it from other unknown coefficients over and over again
in the other cuts. This somehow reduces the efficiency of the
method.

One way to improve the situation is by cutting three propagators
\cite{Bern:2004ky}\cite{Bern:2004bt}. Note that triple cuts where
a single gluon in trapped in between two cut propagators vanish.
This would correspond to a cut in a one-particle channel. In the
next part of this lecture we will reconsider this issue.

Note that the number of scalar boxes with a given triple cut is
less than that with a given unitarity cut. However, in general one
still has to apply reduction techniques. A class of amplitudes for
which triple cuts are very suitable are next-to-MHV (NMHV)
amplitudes \cite{Bern:2004bt}. But, one might expect that this
procedure becomes cumbersome already for NNMHV amplitudes.

It turns out that there is a way of avoiding the reduction
techniques as well as the recalculation of known coefficients.
This is achieved by considering quadruple cuts
\cite{Britto:2004nc} which we now discuss.

\subsubsection{Quadruple Cuts}

Consider a scalar one-loop Feynman integral, $I$. The integral $I$
is a function of the kinematical invariants constructed out of the
external momenta. In general, $I$ is a complicated multi-valued
function with branch cuts that are like domain walls in the space
of kinematical invariants, $\Sigma$. As it is well known, cutting
two propagators in the loop computes the imaginary part of the
integral in a certain region of $\Sigma$. This imaginary part of
$I$ can be thought of as the discontinuity of $I$ across the
branch cut of interest.

Now consider unitarity cuts in several possible channels. One can
ask what is the discontinuity across the intersection of two or
more cuts. The answer is given by the union of the set of cut
propagators! Of particular interest to us is the meaning of
cutting all propagators in a one-loop integral; such a cut
integral computes the discontinuity across the singularity of
highest codimension, which is known as the leading singularity.
For a more extensive discussion and references see
\cite{Eden:1966}\footnote{In \cite{Eden:1966}, the arguments are
made for a massive scalar field theory. However, it turns out that
the relevant results for our discussion can be used in massless
theories with little modifications.}.

As mentioned above, ${\cal N}=4$ one-loop amplitudes of gluons can
be written as a linear combination of scalar box integrals with
rational coefficients. The scalar box integrals can be thought of
as a ``basis of vectors" in some sort of vector space. The idea is
that this basis is in some appropriate sense
orthogonal\footnote{To push the analogy even further, one can
think of the scalar box functions defined in \cite{Bern:1994zx},
which are scalar box integrals nicely normalized, as an
orthonormal basis!} (In less supersymmetric theories there also
are bubble and triangle integrals which break the orthogonality
condition).

The one-loop amplitude $\CA_n^{\rm 1-loop}$ can now be interpreted
as a general vector which can be written as a linear combination
of the basis. All we need is the appropriate way of projecting the
``vector" $\CA_n^{\rm 1-loop}$ onto a given vector $I$ in order to
compute the corresponding coefficient.

{}From our discussion, it is clear that the natural way of doing
this is to consider $\CA_n^{\rm 1-loop}$ in the region near the
leading singularity of $I$, which is unique to $I$. The
discontinuity of $\CA_n^{\rm 1-loop}$ across such a singularity is
the coefficient of $I$, up to a normalization, which is the analog
of the norm of $I$.

Let us see how this works in practice. Recall that the scalar box
integral \rf{ifou} is,
\eqn{ifout}{I_{(K_1,K_2,K_3)} = \int d^4\ell { 1\over
(\ell^2+i\epsilon) ((\ell-K_1)^2 + i\epsilon )((\ell -K_1-K_2)^2 +
i\epsilon )((\ell+K_4)^2 + i\epsilon )}.}
In the expansion of $\CA^{\rm 1-loop}_n$, each $K_i$ in \rf{ifout}
is the sum of the momenta of consecutive external gluons.

The discontinuity across the leading singularity $\Delta_{LS}$ is
computed by cutting all four propagators. This is called a
quadruple cut:
\eqn{wpuvv}{\Delta_{LS} I_{(K_1,K_2,K_3)} = \int d^4\ell
~\delta^{(+)}(\ell^2) ~\delta^{(+)}((\ell-K_1)^2)
~\delta^{(+)}((\ell -K_1-K_2)^2)~\delta^{(+)}((\ell+K_4)^2).}

In order to make the discussion more explicit we introduce
notation for the coefficients of $I$ in the expansion of
$\CA_n^{\rm 1-loop}$ as follows:
 \eqn{ampli}{ \CA_n^{\rm 1-loop} =
\sum_{1<i<j<k<m<n} B_{ijkm} I_{(p_{i+1}+\ldots + p_{j},
p_{j+1}+\ldots +p_{k}, p_{k+1}+\ldots +p_{m})}, }
where the coefficients $B_{ijkm}$ are rational functions of the
spinor products, as mentioned above.

\begin{figure}
\centering
\includegraphics[height=2.in]{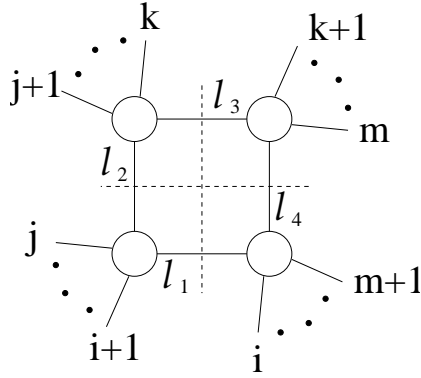}
\caption{A quadruple cut diagram. Momenta in the cut propagators
flows clockwise and external momenta are taken outgoing.  The
tree-level amplitude $\CA^{\rm tree}_1$, for example, has external
momenta $(-\ell_1, i+1,...,j,\ell_2)$.} \label{druple}
\end{figure}

By cutting four given propagators in all possible Feynman diagrams
contributing to $\CA_n^{\rm 1-loop}$ one finds the product of four
tree-level amplitudes integrated over the Lorentz invariant phase
space of four null vectors.

Comparing the two sides of \rf{ampli} we find
\eqn{kasa}{\int d\mu \CA^{\rm tree}_1\CA^{\rm tree}_2\CA^{\rm
tree}_3\CA^{\rm tree}_4 = B_{ijkm} \int d^4\ell
~\delta^{(+)}(\ell^2) ~\delta^{(+)}((\ell-K_1)^2)
~\delta^{(+)}((\ell -K_1-K_2)^2)~\delta^{(+)}((\ell+K_4)^2)}
where the measure is the same one both sides of the integrals,
\eqn{qeqi}{d\mu = d^4\ell ~\delta^{(+)}(\ell^2)
~\delta^{(+)}((\ell-K_1)^2) ~\delta^{(+)}((\ell
-K_1-K_2)^2)~\delta^{(+)}((\ell+K_4)^2),}
and the tree-level amplitudes are defined as follows (see figure
\rf{druple})
\eqn{treeam}{\begin{array}{rlrl} \CA_1^{\rm tree} = &
\CA(-\ell_1,i+1,i+2,\ldots ,j-1,j,\ell_2 ), & \quad \CA_2^{\rm
tree} = & \CA(-\ell_2 , j+1,j+2,\ldots ,k-1, k, \ell_3), \\
\CA_3^{\rm tree} = & \CA(-\ell_3, k+1,k+2,\ldots , m-1,m,\ell_4),
& \quad \CA_4^{\rm tree} = & \CA(-\ell_4, m+1,m+2,\ldots ,i-1,i,
\ell_1).
\end{array} }

In general one might expect that four delta functions localize the
integral producing a Jacobian which is common to both sides of
\rf{kasa} and cancels out to give
\eqn{cool}{ B_{ijkl} = {1\over |{\cal S}|}\sum_{\cal S}\CA^{\rm
tree}_1\CA^{\rm tree}_2\CA^{\rm tree}_3\CA^{\rm tree}_4. }
Here ${\cal S}$ is the set of solutions to the conditions imposed
by the delta functions, and $|{\cal S} |$ is the number of
solutions.

The derivation of the formula for the coefficients \rf{cool}
assumes that the Jacobian is a smooth function and that it does
not vanish for generic momenta of the external gluons and that it
is the same for all solutions ${\cal S}$ \cite{Britto:2004nj}.

It turns out that both assumptions are not valid if at least one
of the momenta $K_i$ in the box integral is null, i.e., if
$K_i^2=0$ for some $i$. This is where the problem of defining a
cut in a one-particle channel, which we mentioned in the
discussion of triple cuts, comes back again.

It is not difficult to see that by using $\delta^{(+)}(\ell^2)$ to
reduce the integration over arbitrary $\ell$'s to those lying in
the future light-cone and by using that, say, $K_1^2=0$ one finds
that two of the three remaining delta functions are enough to
localize the integral. The last delta function, that can be
thought of as part of the Jacobian, imposes an extra constraint on
the external momenta beyond momentum conservation. Therefore, this
makes our two assumptions fail.

This problem of defining a cut in a one-particle channel is the
familiar statement that a gluon cannot decay into two gluons. In
other words, the tree-level amplitude $\CA_1^{\rm tree}$ in
\rf{kasa} vanishes.

It turns out that the way out of both problems is the same.
Consider a Wick rotation of \rf{kasa} in to $--++$
signature\footnote{In some sense it is even more natural to
complexify all momenta.}. In this case one needs all four delta
functions in order to localize the integral. The integration can
be done and produces a smooth and generically nonzero Jacobian.
The reason for this will be clear shortly.

It remains to see what happens to $\CA_1^{\rm tree}$. If it is
still zero it would imply that all coefficients with $K_1^2=0$ are
zero. This is known to be false in MHV and NMHV amplitudes.

Let us look more closely at the tree-level amplitude $\CA_1$ and
the delta function containing only $K_1$. There are two cases,
\eqalign{cases}{\CA_1(K_1^+,\ell^+,(\ell-K_1)^-) &=& {[K_1,~
\ell]^3\over [K_1,~\ell-K_1][\ell-K_1, ~\ell]},\\
\CA_2(K_1^-,\ell^-,(\ell-K_1)^+) &=& {\vev{K_1,~ \ell}^3\over
\vev{K_1,~\ell-K_1}\vev{\ell-K_1, ~\ell}}.}
The delta function is given by $\delta((\ell-K_1)^2)$. This
implies that $\vev{\ell,~K_1}[\ell,~K_1]=0$. As reviewed in
section 2, in Minkowski space with real momenta $\lambda$ and
$\tilde\lambda$ are complex but not independent, i.e.,
$\tilde\lambda = \pm\bar\lambda$ and therefore the only solution
is $\vev{\ell,~K_1} = [\ell,~K_1] = 0$. On the other hand, in
$--++$ signature, $\lambda$ and $\tilde\lambda$ are real and
independent, therefore we can have $\vev{\ell,~K_1} = 0$ while
$[\ell,~K_1] \neq 0$ or vice versa. This is also the reason why
four delta functions are required to localize the integral
\rf{wpuvv}.

This explains how the problem is completely solved. When summing
over the set of solutions ${\cal S}$, one must take into account
the two possibilities, $\vev{\ell,~K_1} = 0$ or $[\ell,~K_1] = 0$.
One of them makes $\CA_1^{\rm tree}$ vanish while the other does
not. Actually, the presence of two solutions is important even in
the case when no $K_i$ is a null vector. The reason is that each
solution produces a function with a square root. However, the
coefficient must be a rational function. The resolution to this
little puzzle is that by adding the two solutions, which differ
only in the branch the square root takes, one always produces a
rational function.

One can easily see that in general there are only two solutions to
the delta function constraints. The two solutions can be found
explicitly in full generality; we refer the reader to
\cite{Britto:2004nj} for the actual formula. This implies that
$|{\cal S}| = 2$. Using this in \rf{cool} we find a formula for
all one-loop ${\cal N} =4$ amplitude coefficients in terms of
tree-level amplitudes,
\eqn{fine}{ B_{ijkl} = {1\over 2}\sum_{h,\cal S}\CA^{\rm
tree}_1\CA^{\rm tree}_2\CA^{\rm tree}_3\CA^{\rm tree}_4.}
The sum on the right hand side of \rf{fine} is over the two
solutions ${\cal S}$ and over all internal particles in the ${\cal
N}=4$ supermultiplet\footnote{This method has been generalized to
one-loop amplitudes in $\CN=8$ supergravity in
\cite{Bern:2005bb,Bjerrum-Bohr:2005xx}.}.

\subsubsection{Examples}

As a simple example consider the coefficient of
$I_{(3+4,5+6,7+1)}$ in $\CA(1^-,2^-,3^-,4^+,5^+,6^+,7^+)$.

In this case, only one internal helicity configuration gives a non
zero contribution and it allows only gluons to run in the loop.

\begin{figure}
\centering
\includegraphics[height=2.in]{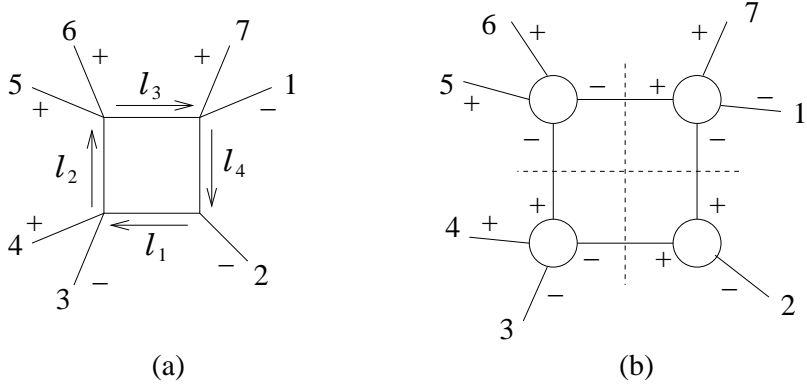}
\caption{$(a)$ Scalar Box Integral $I_{3+4,5+6,7+1}$. $(b)$
Quadruple cut diagram of $\CA_7^{\rm 1-loop}$ corresponding to
$I_{3+4,5+6,7+1}$. Blobs represent tree-level amplitudes.}
\label{threemass}
\end{figure}

Using \rf{fine} we find
\eqn{exco}{
 B_{3572} = \half { [\ell_1 ~\ell_4]^3\over
[\ell_1~2][2~\ell_4] }   { [4~\ell_2]^3 \over
[\ell_2~\ell_1][\ell_1~3][3~4] }   {[5~6]^3\over [6~\ell_3]
[\ell_3~\ell_2] [\ell_2 ~ 5] }   {[\ell_3 ~ 7]^3 \over [7~1] [1~
\ell_4 ] [\ell_4 ~ \ell_3 ] } }

After solving the equations for $\ell_i$ and plugging in the
answer in \rf{exco} one finds a simple expression for $B_{3572}$
\cite{Britto:2004nj, Bern:2004ky}
\eqn{anwi}{-{ \vev{1~2}^3 \vev{2~3}^3 [5~6]^3 \over \vev{7~1}
\vev{3~4} \gb{2 | 3+4| 5} \gb{ 2| 7+1|6} ( \vev{7~1} \gb{ 2| 3+4
|1} - t_2^{[3]} \vev{7~2}) ( t_7^{[2]}\vev{2~4} -\vev{3~4} \gb{
2|7+1| 3})}.}

\pagebreak
\begin{center}
{\bf Acknowledgements}
\end{center}

It is pleasure to thank R. Britto, B. Feng for proof-reading parts
of the manuscript and especially to E. Witten for giving us a
permission to base sections 2-4 of the lecture notes on his
lectures given at PITP, IAS Summer 2004. Work of F. Cachazo was
supported in part by the Martin A. and Helen Chooljian Membership
at the Institute for Advanced Study and by DOE grant
DE-FG02-90ER40542 and that of P. Svr\v{c}ek in part by Princeton
University Centennial Fellowship and by NSF grants PHY-9802484 and
PHY-0243680. Opinions and conclusions expressed here are those of
the authors and do not necessarily reflect the views of funding
agencies.

\end{document}